\documentclass[twocolumn,showpacs,showkeys,preprintnumbers,amsmath,amssymb,nofootinbib]{revtex4}

\usepackage{dcolumn}

\usepackage{amsmath,amsfonts,bbm, graphicx,color}
\usepackage{perpage}
\MakePerPage{footnote}
\usepackage{amssymb}
\usepackage{appendix}
\usepackage[colorlinks]{hyperref}
\usepackage[figure,table]{hypcap}
\usepackage[textsize=tiny]{todonotes}

\hypersetup{
	bookmarksnumbered,
	pdfstartview={FitH},
	citecolor={darkgreen},
	linkcolor={darkblue},
	urlcolor={darkblue},
	pdfpagemode={UseOutlines}}
\definecolor{darkgreen}{RGB}{20,100,20}
\definecolor{darkblue}{RGB}{0,0,130}
\definecolor{darkred}{rgb}{.8,0,0}

\def\>{\rangle}
\def\<{\langle}
\def\E{ {\cal E} }

\def\G{ {\cal G} }

\def\U{ {\cal U} }

\def\U{ {\cal U} }

\def\I{ \mathbbm{1} }

\def\>{\rangle}
\def\<{\langle}

\def \be{\begin{equation}}
\def \ee{\end{equation}}
\def \beq{\begin{equation}}
\def \eeq{\end{equation}}
\def \bea{\begin{eqnarray}}
\def \eea{\end{eqnarray}}

\newtheorem{theorem}{Theorem}[section]

\renewcommand{\[}{\begin{equation}}
\renewcommand{\]}{\end{equation}}

\newcommand{\Tr}[1]{\mathrm{Tr}(#1)}

\newcommand{\ket}[1]{|#1\rangle}
\newcommand{\bra}[1]{\langle#1|}
\newcommand{\braket}[2]{\langle#1|#2\rangle}
\newcommand{\cket}[1]{|\!\!|#1\rangle\!\!\rangle}

\newcommand{\cbraket}[2]{\langle\!\!\langle#1|\!\!|#2\rangle\!\!\rangle}
\newcommand{\pro}[2]{|#1\rangle\langle#2|}

\newcommand{\mean}[1]{\langle#1\rangle}
\newcommand{\norm}[1]{|\!|#1|\!|}
\newcommand{\abs}[1]{|#1|}

\newcommand{\PI}{\bra{\psi}\hat{P}_i\ket{\psi}}

\definecolor{darkred}{rgb}{.8,0,0}
\definecolor{magenta}{RGB}{255,0,255}
\definecolor{green}{rgb}{.2,.6,.2}

\begin{document}

\title{Quantum parameter estimation with imperfect reference frames}

\author{Dominik \v{S}afr\'{a}nek}
\email{pmxdd@nottingham.ac.uk}
\affiliation{School of Mathematical Sciences, University of Nottingham, University Park,
Nottingham NG7 2RD, United Kingdom}
\author{Mehdi Ahmadi}
\email{mehdi.ahmadi@nottingham.ac.uk}
\affiliation{School of Mathematical Sciences, University of Nottingham, University Park,
Nottingham NG7 2RD, United Kingdom}
\author{Ivette Fuentes}\thanks{Previously known as Fuentes-Guridi and Fuentes-Schuller.}
\affiliation{School of Mathematical Sciences, University of Nottingham, University Park,
Nottingham NG7 2RD, United Kingdom}

\date{\today}

\begin{abstract}
Quantum metrology studies quantum strategies which enable us to outperform their  classical counterparts. In this framework, the existence of perfect classical reference frames is usually  assumed. However, such ideal reference frames might not always be available. The reference frames required in metrology strategies can either degrade or become misaligned during the estimation process. We investigate how the imperfectness of reference frames leads to noise which in general affects the ultimate precision limits in measurement of physical parameters. Moreover, since quantum parameter estimation can be phrased as a quantum communication protocol between two parties, our results provide deeper insight into quantum communication protocols with misaligned reference frames. 
Our framework allows for the study of general noise on the efficiency of such schemes.
\end{abstract}


\maketitle

\section{Introduction}

In quantum metrology quantum properties such as squeezing and entanglement are employed to improve the
precision with which physical quantities can be measured~\cite{GLM}. Quantum metrological techniques have
been very fruitful in developing new generation of quantum devices that can outperform their classical
counterparts. In particular, the framework of quantum metrology is very useful for measurement of physical quantities that do not have an associated operator in quantum theory such as time, phase, temperature, acceleration, etc. The process of quantum parameter estimation consists of three stages; the preparation of the probe state of the system, feeding the prepared state into the quantum channel which encodes the parameter of interest into the state of the system and finally decoding the parameter by performing a measurement on the system after it has gone through the channel. In order to reduce the statistical error in the estimation of a physical quantity, this whole process needs to be optimised. This is achieved by optimising over both the preparation of the initial state of the system and the measurement of the final state of the system. For a given prepared state, the quantum Cram\'{e}r-Rao bound provides us with the ultimate precision bound on the measurement of the physical quantities in quantum theory.\\

Holevo and Helstrom  laid the foundations of quantum metrology by phrasing the problem in the context of a communication protocol between two parties~\cite{Hol82,Hel76}. In this paradigm Alice chooses a quantum system to encode a message which she then sends to Bob. For example she might choose to use either a spin-$\frac{1}{2}$ system or a quantum harmonic oscillator as carrier of her message. The message can be encoded as a phase parameter in the state of the spin-$\frac{1}{2}$ particle or  in the state of the quantum harmonic oscillator. Bob then performs a measurement on the system in order to decode the message, i.e. the encoded phase parameter. The quality of this communication protocol can be improved by optimisation of both stages of the protocol, namely Alice's encoding process and Bob's decoding process. However, standard approaches to quantum communication, such as encoding qubits into polarisation degree of freedom of photons, require that all parties have knowledge of a shared reference frame. This means that in the absence of such knowledge, the involved parties need to initially establish aligned reference frames.  Despite the considerable amount of progress in the development of protocols for aligning reference frames such as clock synchronisation and Cartesian frame alignment~\cite{SSR-RF}, maintaining aligned reference frames is still a large obstacle in achieving such tasks. For instance when the parties are in relative motion with respect to each other, the relative orientation of their local reference frames can change in time~\cite{Exper}. However, quantum reference frames (QRF) enable us to circumvent this problem. Let us briefly explain what QRFs are and in what way they differ from classical reference frames (CRF).\\

Aharonov and Susskind in their seminal papers \cite{Aha,Sus}, showed that the concept of reference frame
can be suitably accommodated in quantum theory. In recent years, such treatment of reference frames in quantum theory, i.e. as quantum objects  has led to the formalism of ``\textsl{Quantum reference frames}''  \cite{Pop1,SSR-RF}.
A QRF is different from its classical counterpart in two ways. First, due to its quantum nature, it has an inherent uncertainty and the measurement results are only an approximation of what would be obtained using a classical reference frame. Second, each time the QRF is used, it suffers a back-action, which causes the future measurements to be less accurate. Phase measurement of single-rail qubits relative to a QRF has been investigated in~\cite{Terrydeg}, while the degradation of a directional QRF has been analysed in~\cite{Terrydeg,PY,UnitaryDeg}. In the past few years, the ``\textsl{resource theory of quantum reference frames}'', also known as the ``\textsl{resource theory of asymmetry}'', has been developed. This resource theory provides us with a very useful framework wherein the QRFs are the ``\textsl{resource states}''~\cite{RTA1,RTA2,RTA3,RTA4}. They enable us to achieve quantum information processing tasks without first establishing a shared reference frame. In such schemes, a QRF stands in for the possibility of performing tasks in the absence of a common CRF, in the same way that entangled states allow for the possibility of performing non-local quantum operations.\\

In this paper, we bring to bear the powerful machinery of quantum metrology to study the ultimate precision bounds in measurement of physical parameters with respect to QRFs.  First we explain the connection between the quantum mechanical treatment of reference frames and  quantum parameter estimation in the presence of noise. Then we investigate how the ultimate precision in measurement of a parameter decreases due to inaccessibility of a perfect CRF.  In order to do so,  we analyse the decrease in quantum Fisher information as a result of not having access to a perfect reference frame for the physical quantity of interest. In particular, we provide necessary and sufficient conditions for two extreme cases that can occur in quantum parameter estimation with imperfect frames of reference. The first case is when the absence of a perfect reference frame does not affect the precision with which one can measure the parameter and the second case is when measurement of the parameter of interest is no longer possible due to not having access to a CRF. Motivated by this analysis we split the problem into two separate cases. The first case is when the noise caused by the quantum nature of the RF commutes with the channel of interest and the second case is when the channel of interest and the noise do not commute. We show how the non-commutative noise allows for the estimation of the parameter even in the absence of DFSs. Moreover, we explain the connection between noisy quantum metrology and alignment-free quantum communication. In the end, we present three examples in order to further clarify different aspects of quantum metrology with imperfect frames of reference.

The structure of this paper is as follows: In section~\ref{QCR} we briefly review some of the mathematical tools from quantum metrology, in particular Quantum Fisher information (QFI) and symmetric logarithmic derivative (SLD). In section~\ref{AFC} we represent the general scheme of the alignment-free  communication protocols. In section~\ref{QMQRF} we present the general framework for quantum parameter estimation in the absence of an ideal CRF and we discuss its relation to the alignment-free communication protocols. In section~\ref{Ex} we present three examples in which we explain different aspects of quantum parameter estimation in the absence of aligned CRFs. Finally in section~\ref{DO} we discuss the results of this paper and we mention some of our research interests in quantum parameter estimation  as possible future work.

\section{Preliminaries}

\subsection{Quantum Cram\'{e}r-Rao bound}\label{QCR}
\begin{figure}[t]
\includegraphics[width=\linewidth]{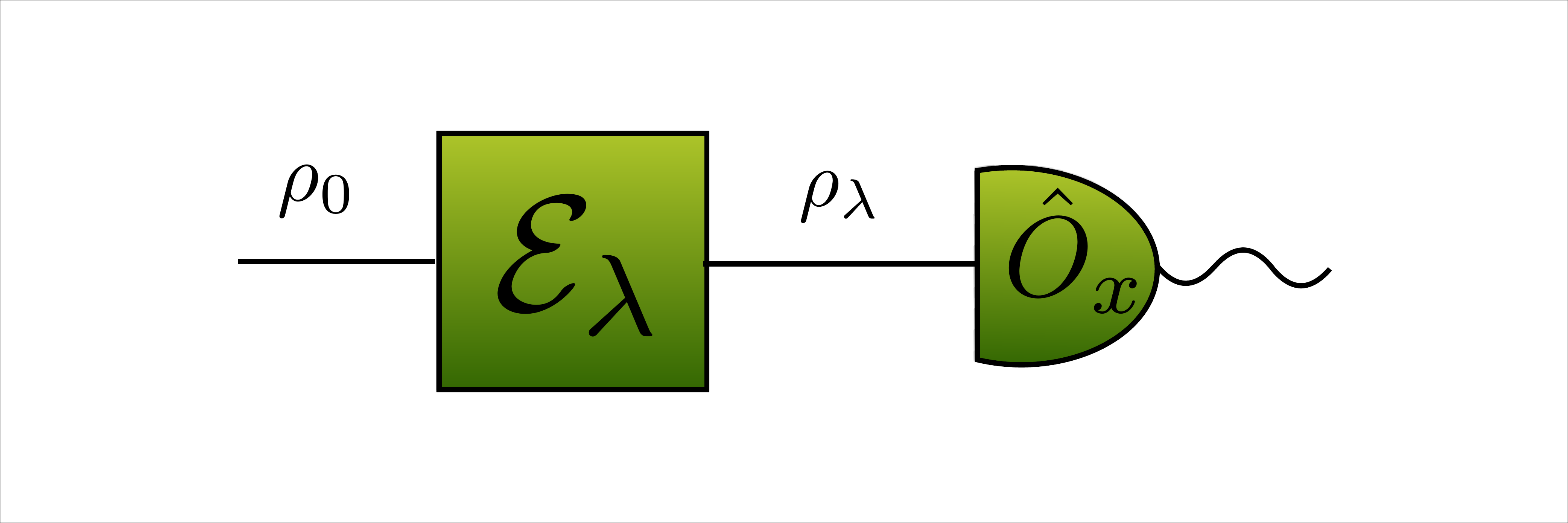}
\caption{Quantum parameter estimation: In order to measure an unknown parameter $\lambda$ of the channel, a probe state is fed into the channel of interest. Then a measurement is performed on the final state of the system which makes it possible to extract information about the parameter of interest.} \label{QCRB}
\end{figure}

The main goal in quantum metrology is to estimate an unknown parameter $\lambda$ of a completely positive trace-preserving (CPTP) quantum channel, $\E_{\lambda}$. In order to do so, first a probe state is prepared which is then fed into the channel of interest. Finally a measurement on the final state of the probe will enable us to estimate $\lambda$, where $\lambda$ is a physical quantity such as time, phase, temperature, acceleration, etc.  Given a measurement strategy, the conditional probability of obtaining outcome $x$ when the initial state is $\rho_{\lambda}$ is given by  $p(x|\lambda)=\Tr{\hat{O}_{x}\rho_{\lambda}}$, where $\{\hat{O}_{x}\}$ are elements of a complete positive-operator valued measure (POVM) corresponding to the chosen measurement strategy (see Fig.\ref{QCRB}). The lower bound on how precise we can estimate $\lambda$ is given by the ``\textsl{classical Cram\'{e}r-Rao bound}'', i.e. $\langle (\Delta \hat{\lambda})^{2}\rangle\geq\frac{1}{NF(\lambda)}$, where the classical Fisher information $F(\rho_{\lambda})$ is defined as

\[\label{CFI}
F(\rho_{\lambda})=\int\!dx \frac{1}{p(x|\lambda)} \left[\frac{d\, p(x|\lambda)}{d\lambda}\right]^{2},
\]
and $N$ is the number of repeated measurements. In other words, classical Fisher informations is an operational measure which tells us how much information we can gain about the unknown parameter $\lambda$ by choosing a certain measurement strategy.\\

Braunstein and Caves showed that optimisation over all the possible quantum measurements provides an even more stringent lower bound~\cite{BraCaves}, i.e.
\be\label{Cramer-Rao}
N\langle (\Delta \hat{\lambda})^{2}\rangle\geq\frac{1}{F(\rho_\lambda)}\geq \frac{1}{H(\rho_\lambda)},
\ee
where $H(\rho_\lambda)$ is the Quantum Fisher information. This quantity is closely related to the Symmetric logarithmic derivative $L({\rho_{\lambda}})$ which is defined by  $2 d\rho_{\lambda}/ d\lambda=L({\rho_{\lambda}}) \rho_{\lambda}+\rho_{\lambda} L({\rho_{\lambda}})\,$. In particular
in the basis  $\{|\psi_i\rangle\}$ in which $\rho_{\lambda}$ is diagonalised, i.e. $\rho_{\lambda}=\sum_i p_i|\psi_i\rangle\langle\psi_i|$, the SLD and the QFI can be written as

\begin{equation}\label{SLD}
L({\rho_{\lambda}})=2\sum_{i,j}\frac{\langle \psi_i|\partial_{\lambda}\rho_{\lambda}|\psi_{j}\rangle}{p_i+p_j} |\psi_{i}\rangle\langle\psi_{j}|,
\end{equation}
\begin{equation}\label{QFI}
H(\rho_{\lambda})=2\sum_{i,j}\frac{\left|\langle \psi_i|\partial_{\lambda}\rho_{\lambda}|\psi_{j}\rangle\right|^2}{p_i+p_j},
\end{equation}
with the relation
\[\label{HSLDrelation}
H(\rho_{\lambda})=\Tr{\partial_\lambda \rho_\lambda L({\rho_{\lambda}})}.
\]

The summations above do not include the terms with $p_i=p_j=0$. The optimal POVM which achieves the quantum Cram\'{e}r-Rao bound \eqref{Cramer-Rao}, can be constructed from the eigenstates of $L(\rho_{\lambda})$~\cite{paris}.\\

Note that in this section the existence of ideal CRFs was assumed. This treatment of reference frames can lead to great deal of confusion. For instance, in \cite{Rafal} the role of an external phase reference frame in interferometric setups has been analysed. In this paper we analyse the decrease in quantum Fisher information as a result of not having access to a perfect reference frame for the physical quantity $\lambda$. In particular, we present necessary and sufficient conditions for two extreme cases; the case where the QFI does not decrease when a CRF is lacking and the case where the QFI vanishes due to imperfectness of the RF, i.e. one can no longer extract the parameter $\lambda$.

\subsection{Alignment-free communication}\label{AFC}
As mentioned earlier, quantum parameter estimation can be phrased as a communication protocol between two parties. In this section we briefly review how QRFs have been employed in order to achieve alignment-free communication protocols~\cite{QCQRF}.\\

Consider $g\in G$ to be the group element that describes the passive transformation from Alice's to Bob's reference frame. Furthermore, since Bob is completely unaware of the relation between his local RF and Alice's local RF, we can assume that the group element $g$ is completely unknown. It follows that if Alice prepares a state $\rho_{A}$ relative to her local reference frame, then relative to Bob's RF this state is seen as\footnote{We will restrict our attention to Lie-groups that are compact, so that they possess a group-invariant (Haar) measure ${\mathrm{d}} g$. We refer the readers for more details to \cite{SSR-RF}.}
\begin{equation}\label{gtwirl}
\rho_{B}={\cal{G}}[\rho_{A}]=\int {\mathrm{d}} g U(g) \rho _{A}U(g)^{\dag}.
\end{equation}

Therefore, lacking such a shared reference frame is equivalent to having a noisy completely positive trace-preserving map which is known as the ``\textsl{g-twirling map}'', i.e. $\G(\rho_{A})$.
However, despite the fact that Alice has no information about the group element $g$ that relates her local RF to Bob's local RF, she can still encode information in the so called ``\textsl{Decoherence-free subsystems (DFS)}''~\cite{DFS1,DFS2,DFS3,SSR-RF}. These subsystems are resilient to the decoherence caused due to the lack of knowledge about the relative direction of the local reference  frames. The efficiency of this protocol depends on the dimensionality of the largest DFS, i.e. the subsystem which possesses the largest number of degenerate eigenstates. Such communication scenarios in the absence of a shared Cartesian reference frame have been analysed before~\cite{SSR-RF}. The idea is to encode logical qubits into rotationally invariant states of multiple physical qubits. In this case the  number of logical qubits per number of physical qubits that can be transmitted scales as $1-M^{-1} \log_{2}(M)$, where $M$ is the number of transmitted physical qubits. This remarkable result proves that in the limit of $M\to\infty$ one logical qubit can be sent per one physical qubit. Therefore in this limit the efficiency of this scheme is the same as the scenario wherein the reference frames are aligned. This protocol has also been studied for the situation in which the parties do not share a common background phase reference frame~\cite{QCQRF,WAY}.\\

\section{Parameter estimation with imperfect reference frames}\label{QMQRF}
As we explained in section \ref{QCR}, the standard scenarios considered in quantum metrology normally presume the existence of perfect classical reference frames. In this section we investigate how the ultimate precision in measurement of a parameter decreases due to lack of access to a perfect CRF.\\

Let us first briefly explain the general picture of the estimation of a parameter in the absence of a perfect external RF. We consider the case where the parameter $\lambda$ is encoded into the fiducial state via a unitary channel $\U_{\lambda}$. After this encoding process, we need to choose the optimal measurement in order to extract the maximum amount of information about $\lambda$. We then need a suitable RF with respect to which we are able to perform the chosen measurement. For instance if we wish to measure time we need a clock or if we need to measure phase we will need a phase reference frame. The absence of such reference frames can be viewed as a noisy quantum channel, i.e.

\begin{equation}\label{Htwirl}
\G[\rho_{\lambda}]=\lim_{T\rightarrow\infty} \frac{1}{T}\int_{0}^{T} dt\ U(t)\rho_{\lambda}U(t)^{\dag},
\end{equation}

where $\rho_{\lambda}=|\psi_{\lambda}\rangle\langle\psi_{\lambda}|$, $|\psi_{\lambda}\rangle=U(\lambda)|\psi_{0}\rangle$, $U(\lambda)=e^{-i\hat{K}\lambda}$ and $U(t)=e^{-i\hat{G}t}$. Note that $\hat{G}$ and $t$ are determined by the type of RF that is lacking. For instance if we need to measure phase of a quantum harmonic oscillator then $t$ is a phase and $\hat{G}=\hat{N}$, where $\hat{N}$ is the total number operator~\cite{SSR-RF}.  \\ 

Assuming the spectral decomposition $\hat{G}=\sum_{i}G_{i}\hat{P}_{i}$, where $\hat{P}_{i}$s are the projectors into subspaces with eigenvalues $G_{i}$ and $\sum_{i}\hat{P}_{i}=\I$, one can easily check that the state $\G[\rho_{\lambda}]$ in~\eqref{Htwirl} can be written as
\begin{equation}\label{rhoB}
\G[\rho_{\lambda}]=\sum_{i}\hat{P}_{i}\rho_{\lambda}\hat{P}_{i}.
\end{equation}

\begin{figure}[t]
\includegraphics[width=\linewidth]{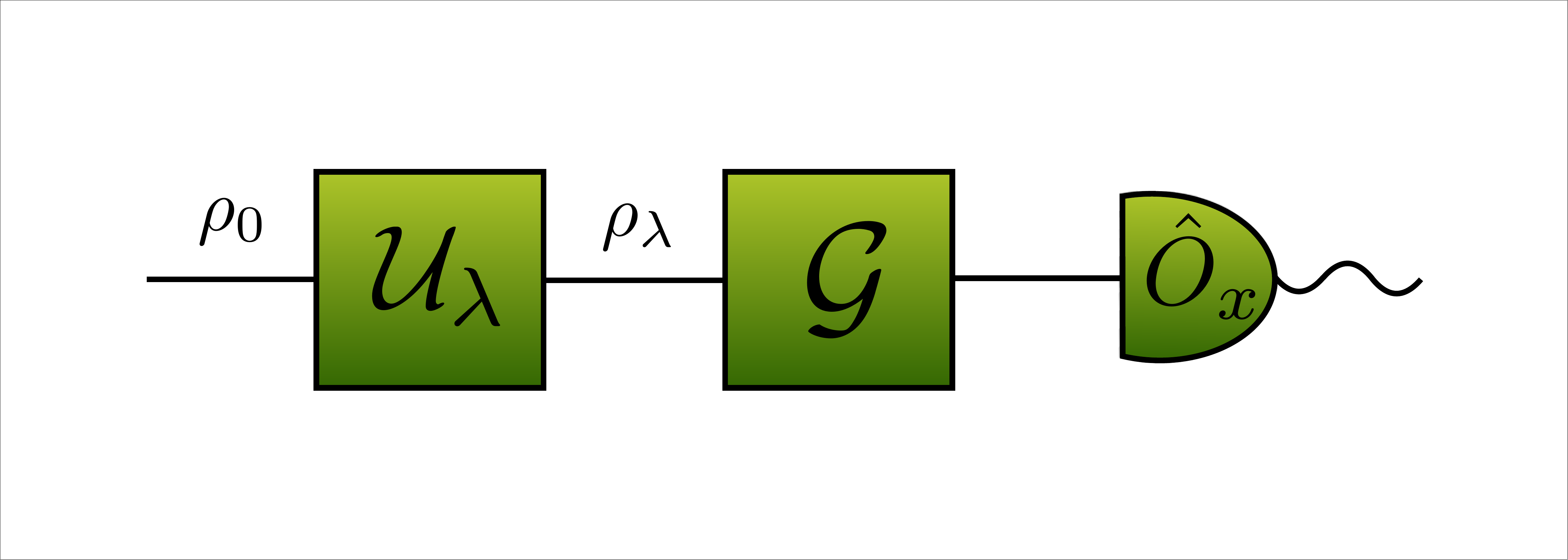}
\caption{Quantum parameter estimation without perfect classical reference frames} \label{NQCRB}
\end{figure}
As depicted on figure \ref{NQCRB} the problem of parameter estimation without a CRF can be phrased as the problem parameter estimation in the presence of noise. Note that in the special case of commuting $\hat{K}$ and $\hat{G}$ operators, i.e. when $[\hat{K},\hat{G}]=0$, this general noise reduces to the well-known ``\textsl{collective dephasing noise}'' with uniform prior probability~\cite{lidarbook}. In fact in the quantum information protocols considered in~\cite{SSR-RF} and the references therein it is assumed that $[\hat{K},\hat{G}]=0$.\\

Here, we define ``\emph{Quantum Fisher information loss}'' as
\begin{equation}\label{QFIlossdef}
l(\rho_{\lambda},\hat{G})=H(\rho_{\lambda})-H(\G[\rho_{\lambda}]).
\end{equation}
This operational measure enables us to analyse how much information is lost if instead of ideal CRFs we only have access to imperfect frames of reference. We investigate the decrease in the accuracy of measurements in both cases of commuting and non-commuting $\hat{K}$ and $\hat{G}$.

\subsection{General framework}
As mentioned earlier we restrict our analysis to pure initial states of the system, i.e.\ $\rho_{\lambda}=\pro{\psi_{\lambda}}{\psi_{\lambda}}$. Using Eq.~\eqref{QFI} for QFI, the properties of the quantum channel $\G$ \eqref{rhoB} and the Parseval identity (for more details see appendix \ref{FirstFormula}), we derive Bob's QFI as
\[\label{HrhoB}
H(\G[\rho_{\lambda}])=4\braket{\partial_\lambda\psi_{\lambda}}{\partial_\lambda\psi_{\lambda}}-
4\sum_i\frac{(\mathfrak{Im}\bra{\psi_{\lambda}}\hat{P}_i\ket{\partial_\lambda\psi_{\lambda}})^2}{\bra{\psi_{\lambda}}\hat{P}_i\ket{\psi_{\lambda}}},
\]
where the summation is over the indices $i$ for which $p_i=\bra{\psi_{\lambda}}\hat{P}_i\ket{\psi_{\lambda}}\neq0$. Note that we will use this convention throughout the rest of the paper.\\

Differentiating $\braket{\psi_{\lambda}}{\psi_{\lambda}}=1$ yields
\[\label{purelyimaginary}
\braket{\psi_{\lambda}}{\partial_\lambda\psi_{\lambda}}+\braket{\partial_\lambda\psi_{\lambda}}{\psi_{\lambda}}=2\mathfrak{Re}\braket{\psi_{\lambda}}{\partial_\lambda\psi_{\lambda}}=0,
\]
and therefore $\braket{\psi_{\lambda}}{\partial_\lambda\psi_{\lambda}}$ is purely imaginary. The fact that $\braket{\psi_{\lambda}}{\partial_\lambda\psi}$ is purely imaginary will be used frequently in the rest of the paper.\\

 Note that if we trivially choose $\sum_i\hat{P}_i\equiv \hat{P}_1=\I$, i.e. in the presence of a shared reference frame, we recover the result for the Quantum Fisher information of pure states
\[\label{paris}
H(\rho)=4\braket{\partial_\lambda\psi_{\lambda}}{\partial_\lambda\psi_{\lambda}}-
4|\braket{\psi_{\lambda}}{\partial_\lambda\psi_{\lambda}}|^2,
\]
as was proved in \cite{paris}.\\

Using Eqs. \eqref{HrhoB} and \eqref{paris}, we find the expression below for the  \emph{Quantum Fisher information loss} as defined by \eqref{QFIlossdef}
\[\label{QFIL}
l(\rho,\hat{G})=4\Big(\sum_i\frac{(\mathfrak{Im}\bra{\psi_{\lambda}}\hat{P}_i\ket{\partial_\lambda\psi_{\lambda}})^2}{\bra{\psi_{\lambda}}\hat{P}_i\ket{\psi_{\lambda}}}-|\braket{\psi_{\lambda}}{\partial_\lambda\psi_{\lambda}}|^2\Big).
\]
This quantity is always non-negative as expected (See appendix \ref{nonnegQFIl}), it simply means that the accuracy with which one can measure $\lambda$ in presence of a perfect CRF can not be less than the accuracy with which he/she can measure $\lambda$ in the absence of such RFs. In the theorem below we formalise the necessary and sufficient conditions for two extreme cases; the first case is where the precision in measurement of $\lambda$ remains the same both in the absence or the presence of a  perfect CRF and the second case is where the measurement of $\lambda$ is not possible anymore due to inaccessibility of such reference frames. Note that from this point on we drop the subscript $\lambda$.
\begin{theorem}\label{theorem}
$0\leq l(\rho,\hat{G})\leq H(\rho)$.

$l(\rho,\hat{G})=0$ (no loss) $\Leftrightarrow$ There exists a complex number $c$ such that
\[
\ket{\widetilde{\partial_\lambda\psi}}:=\sum_i\frac{\mathfrak{Im}\bra{\psi}\hat{P}_i\ket{\partial_\lambda\psi}}{\langle\psi|\hat{P}_i|\psi\rangle}\hat{P}_i\ket{\psi}=c\ket{\psi}
\]
or equivalently
\[\label{cnoloss}
\exists c\in\mathbb{C},\ \forall i,\ \mathfrak{Im}\bra{\psi}\hat{P}_i\ket{\partial_\lambda\psi}=c\PI.
\]

$l(\rho,\hat{G})=H(\rho)$ (max loss) $\Leftrightarrow$
\[
\braket{\widetilde{\partial_\lambda\psi}}{\widetilde{\partial_\lambda\psi}}=\braket{{\partial_\lambda\psi}}{{\partial_\lambda\psi}}
\]
or equivalently
\[\label{maxlosstheorem}
\forall i,\ \mathfrak{Re}\bra{\psi}\hat{P}_i\ket{\partial_\lambda\psi}=0\ \wedge\ \forall \ket{\phi_j},\ \braket{\phi_j}{\partial_\lambda\psi}=0,
\]
where $\{\frac{\hat{P}_i\ket{\psi}}{\sqrt{p_i}},\ket{\phi_j}\}_{i,j}$ is an orthonormal basis of the Hilbert space.
Moreover, Quantum Fisher information loss can be written as

\[
l(\rho,\hat{G})=4\braket{\widetilde{\partial_\lambda\psi}}{\widetilde{\partial_\lambda\psi}}-4\abs{\braket{\psi}{\widetilde{\partial_\lambda\psi}}}^2.
\]

\end{theorem}

Let us add three notes to this theorem. First, after summing over all the indices $i$ in Eq.~\eqref{cnoloss} and using Eq.~\eqref{purelyimaginary}, one can easily find that $c=\mathfrak{Im}\braket{\psi}{\partial_\lambda\psi}=-i\braket{\psi}{\partial_\lambda\psi}$. Second, without loss of generality in Eq.~\eqref{cnoloss} we can restrict our analysis to the terms for which  $p_i=\PI\neq0$, since using Schwarz inequality it can be checked that the condition~\eqref{cnoloss} holds trivially if $p_i=0$. Third, the set of states $\{\ket{\phi_j}\}_j$ are orthonormal states which together with the set of normalised states $\{\frac{\hat{P}_i\ket{\psi}}{\sqrt{p_i}}\}_i$ make a complete set. We can always find the set of states $\{\ket{\phi_j}\}_j$ via the Gram-Schmidt process for orthonormalisation of a set of vectors. Alternatively, we can see them as eigenvectors of $\G(\rho)$ with the respective eigenvalue $0$.\\

Using similar analysis we can find the SLD operator in \eqref{SLD} as
\[\label{NSLD}
L(\G(\rho))=\sum_i\pro{\varphi_i}{\psi_i}+\pro{\psi_i}{\varphi_i},
\]

where $|\psi_{i}\rangle$ and $|\varphi_{i}\rangle$ are defined as
\begin{equation}
\begin{split}
\ket{\psi_i}&=\frac{\hat{P}_i\ket{\psi}}{\sqrt{p_i}}\\
\ket{\varphi_i}&=\frac{1}{\sqrt{p_i}}\left(2\hat{P}_i\ket{\partial_\lambda\psi}-
\langle{\psi_i}\ket{\partial_\lambda\psi}\ket{\psi_i}\right).
\end{split}
\end{equation}

Now we can use this SLD operator whenever we lack a perfect CRF in order to find the POVM that can optimally distinguish between the two neighbouring states $\rho_{\lambda}$ and $\rho_{\lambda+\delta\lambda}$, where $\delta\lambda$ is an infinitesimal increment in the parameter $\lambda$.\\

So far we have analysed the problem in hand by focusing on the projectors $\hat{P}_i$. Since these projectors are constructed from the eigenvectors of the operator $\hat{G}$, we can instead write every derived expression in terms of these eigenvectors (See appendix \ref{alternative}).

\subsection{Analysis of commutative and non-commutative noise due to lacking a perfect CRF}\label{correlation}
In this section we analyse QFI in terms of the hermitian operator $\hat{K}$ which imprints the parameter $\lambda$ into the fiducial state $|\psi_0\rangle$ and $\hat{G}$ which is the generator of the noisy channel. This way we split the problem into two different cases. The first case is where the encoding process in general does not commute with the noisy channel, i.e. $[\hat{K},\hat{G}]\neq0$. We call such noise non-commutative. The second is when when the noise is commutative\footnote{If the noise is commutative, it simply means that the noisy channel \eqref{Htwirl} commutes with the encoding process. In that case our results can be also applied on systems where the noise \eqref{rhoB} precedes the encoding operation $U(\lambda)$, or more specifically, systems with mixed fiducial state $\rho_0$.}, i.e. $[\hat{K},\hat{G}]=0$. For commutative noise formulas usually simplify and are easier to interpret.

Using Eq.~\eqref{HrhoB} we derive an alternative form for the Quantum Fisher information in the absence of a perfect RF as
\[\label{noncommutingKG}
H(\G[\rho])=4\bra{\psi}\hat{K}^2\ket{\psi}-
\sum_i\frac{\bra{\psi}\{\hat{P}_i,\hat{K}\}\ket{\psi}^2}{\bra{\psi}\hat{P}_{i}\ket{\psi}},
\]
where $\{\cdot\ \!,\cdot\}$ denotes anti-commutator.  Noting that $\hat{G}$ commutes with $\hat{K}$ if and only if all the projectors $\hat{P}_i$ commute with $\hat{K}$, for the case of commuting $\hat{G}$ and $\hat{K}$ Eq.~\eqref{noncommutingKG} reduces to
\[\label{commutingKG}
H(\G[\rho])=4\bra{\psi_0}\hat{K}^2\ket{\psi_0}-
4\sum_i\frac{\bra{\psi_0}\hat{P}_{i}\hat{K}\ket{\psi_0}^2}{\bra{\psi_0}\hat{P}_{i}\ket{\psi_0}}.
\]
Now let us revisit the no-loss and maximum-loss conditions that we presented in theorem \ref{theorem}. These conditions can be written in terms of projectors $\hat{P}_i$ and generator $\hat{K}$ as
\[\label{nolosscondition}
l(\rho,\hat{G})=0\ \Leftrightarrow\ \forall i,\ \mean{\{\hat{P}_i,\hat{K}\}}_\rho=2\mean{\hat{K}}_\rho\mean{\hat{P_i}}_\rho.
\]
\begin{equation}\label{maxloss}
\begin{split}
\ \!\!\!\!\!\!\!\!\!\!\!\!\!\!\!l(\rho,\hat{G})=H(\rho)\ \Leftrightarrow\ &\forall i,\ \mean{[\hat{P_i},\hat{K}]}_\rho=0\\
&\wedge\ \forall \ket{\phi_j},\ \bra{\phi_j}\hat{K}\ket{\psi}=0,
\end{split}
\end{equation}

where by $\mean{\cdot}_{\rho}$ is the expectation value with respect to state $\rho$.
If we assume that $[\hat{K},\hat{G}]=0$ and that the operator $\hat{G}$ has a non-degenerate spectrum, i.e. all the projectors $\hat{P_i}$ are rank-$1$ projections, then in the absence of a perfect CRF all the information about $\lambda$ will be lost  (For details see appendix \ref{alternative}).
This is no longer the case when the two operators do not commute. This means that, even though the decoherence-free subspaces are crucial for successful encoding of parameter $\lambda$ in the commuting case, such subspaces are not necessary in the non-commuting case. As we will explain in section \ref{RAFC}, this fact stands in for the possibility of alignment-free communication whenever the spectrum of $\hat{G}$ is non-degenerate. We will present examples of these two different cases, i.e.  commutative and non-commutative noise in section \ref{Ex}.\\

We can write the no-loss condition \eqref{nolosscondition} in a more intuitive way as
\[
l(\rho,\hat{G})=0\ \Leftrightarrow\ \forall i,\ \mathrm{Cov}_\rho(\hat{P}_i,\hat{K})=0,
\]

where the covariance\footnote{ Here we use the symmetrised form of covariance. For other forms of covariance see~\cite{SymCov}.} of two observables $\hat{A}$ and $\hat{B}$ is defined as
$\mathrm{Cov}_\rho(\hat{A},\hat{B})=\frac{1}{2}\mean{\{\hat{A}-\mean{\hat{A}},\hat{B}-\mean{\hat{B}}\}}_\rho
=\frac{1}{2}\mean{\{\hat{A},\hat{B}\}}_\rho-\mean{\hat{A}}_\rho\mean{\hat{B}}_\rho$, and the variance can be written as $\mathrm{Var}_\rho(\hat{A})=\mathrm{Cov}_\rho(\hat{A},\hat{A})$. Covariance is a measure of correlations between  two observables $\hat{A}$ and $\hat{B}$ with respect to the state $\rho$\footnote{ We refer the readers to~\cite{Luo} for details of the relation between the correlations of two observables and the covariance of observables.}. Multiplying this equation by the eigenvalues $G_i$ and summing over all the indices $i$, we can write the necessary condition for not loosing any information about $\lambda$ as
\[\label{noloss}
l(\rho,\hat{G})=0\ \Rightarrow\ \mathrm{Cov}_\rho(\hat{G},\hat{K})=0.
\]
This means that if operators $\hat{K}$ and $\hat{G}$ are correlated with respect to the pure initial state $\rho$, i.e. $\mathrm{Cov}_\rho(\hat{G},\hat{K})\neq0$, then some information is lost due to the imperfectness of RF, i.e. $l(\rho,\hat{G}) > 0$.

It is worth emphasising that this condition is not sufficient. As an example consider the operators $\hat{K}=\pro{2}{2}$, $\hat{G}=6\pro{0}{0}+3\pro{1}{1}+4\pro{2}{2}$, and the fiducial state $\ket{\psi_0}=\frac{1}{\sqrt{6}}\ket{0}+\frac{1}{\sqrt{3}}\ket{1}+\frac{1}{\sqrt{2}}\ket{2}$. In this example the covariance between $\hat{G}$ and $\hat{K}$  is zero, nevertheless, since $\hat{K}$ and $\hat{G}$ commute and the fact that no non-degenerate subspace exists, we will not be able to extract any information about $\lambda$. \\ 

Similar to the no-loss condition in \eqref{noloss}, using Eq.~(\ref{maxloss}) the necessary condition for the extreme case of loosing all the information can be written as
\[\label{maxlossKG}
l(\rho,\hat{G})=H(\rho)\ \Rightarrow\ \mean{[\hat{G},\hat{K}]}_\rho=0.
\]
This means that if for a given initial state $\mean{[\hat{G},\hat{K}]}_\rho$ is non-zero, there is the possibility of extracting some information about parameter $\lambda$ even in the absence of a CRF.\\

The Quantum Fisher information of a unitary channel in the absence of any noise can be computed from Eq.~$\eqref{paris}$ or  alternatively by~\cite{paris}
\[\label{QFIunitary}
H_U(\rho)=4\big(\mean{\hat{K}^2}_\rho-\mean{\hat{K}}_\rho^2\big)=4\mathrm{Var}_\rho(\hat{K}),
\]
where $\hat{K}$ is again the generator of the unitary channel.
We presented the Quantum Fisher information loss due to the quantum nature of the RF in \eqref{QFIL}, we can re-write this equation in terms of the projectors $\hat{P}_{i}$ and the generator $\hat{K}$ as
\[\label{losscovformula}
l(\rho,\hat{G})=4\sum_i\frac{(\mathrm{Cov}_\rho(\hat{P_i},\hat{K}))^2}{p_i}.
\]
This enables us to write the Quantum Fisher information in the absence of a perfect CRF in a form which is easier to compare to the quantum information in the presence of classical frames of reference given in Eq.~\eqref{QFIunitary}. We present this form of QFI in the following theorem.

\begin{theorem}\label{nicetheorem}
For a fiducial pure state $|\psi_0\rangle$, a generator $\hat{K}$ of a unitary operator $U(\lambda)=\text{exp}(-i\hat{K}\lambda)$ and projectors $\hat{P}_i$ of the g-twirling map in \eqref{rhoB}, the Quantum Fisher information of the state $\G[\rho_{\lambda}]$ is
\begin{equation}\label{BobQFI}
\begin{split}
 H(\G[\rho_{\lambda}])
&=4\mathrm{Var}_{\rho_{\lambda}}(\hat{K})-4\sum_i
p_i\left[\mathrm{Cov}_{\rho_{\lambda}}\left(\frac{\hat{P}_i}{p_i},\hat{K}\right)\right]^2,
\end{split}
\end{equation}
where $\rho_{\lambda}=|\psi_{\lambda}\rangle\langle\psi_{\lambda}|$, $\ket{\psi_{\lambda}}=U(\lambda)\ket{\psi_0}$ and $p_i=\mean{\hat{P_i}}_{\rho_\lambda}$.
\end{theorem}

From Eq.~\eqref{BobQFI} we deduce that the decrease in QFI is proportional to the mean of squared covariances between the normalized\footnote{$\mean{\hat{P}_i/{p_i}}_{\rho}=1$} projectors ${\hat{P}_i}/{p_i}$ and the encoding operator $\hat{K}$. This means that  the more projectors ${\hat{P}_i}/{p_i}$ are correlated with the encoding operator $\hat{K}$, the more precision is lost. Roughly speaking, in order to lose the minimum amount of precision one should choose an encoding operator $\hat{K}$ which is less correlated with the decoherence caused by the noisy channel $\G$. For an explicit example we refer the readers to the third example of section~\ref{Ex}.\\

For the special case of a commutative noise, i.e. when $\hat{K}$ and $\hat{G}$ commute, the expression \eqref{BobQFI} can be further simplified. In this case we have $\text{Cov}(\hat{P}_i,\hat{K})=p_i(\mean{\hat{K}}_{\rho_i}-\mean{\hat{K}}_{\rho})$, where $\rho_i=\frac{\hat{P}_i\rho_{\lambda}\hat{P}_i}{p_i}$. This causes the QFI in the absence of a perfect RF to reduce to $H(\G[\rho_{\lambda}])=\sum_i p_i H_U(\rho_i)$, where $H_U$ is the QFI for the unitary channel as given in Eq.~\eqref{QFIunitary}. 
For an explicit example, we refer the readers to the first example of section \ref{Ex}.

\subsection{Relation to alignment-free communication protocols}\label{RAFC}
As pointed out in~\cite{SSR-RF}, the problem of communication between two parties who do not share a common CRF can be mapped into the problem of communication between the parties via the noisy quantum channel in \eqref{rhoB}, while assuming that their local reference frames are aligned. As an example consider the case where Alice and Bob do not share a Cartesian reference frame as depicted in figure $3a$. Suppose Alice encodes a parameter $\lambda$ in
a qubit plus a quantum sample of her local Cartesian RF, i.e. a quantum Cartesian RF. Also assume that the encoding process is done via a unitary channel $U(\lambda)=\text{exp}(-i\hat{K}\lambda)$, where $\hat{K}$ is the generator of the unitary transformation. She then transmits the qubit together with the quantum token of her local Cartesian reference frame, namely the state $|\psi_{\lambda}\rangle=U(\lambda)|\psi_q\rangle\otimes|\psi_{QRF}\rangle$, where $|\psi_{q}\rangle$ and $|\psi_{QRF}\rangle$ are the initial state of qubit and the QRF respectively.\\

The state of the whole system will be decohered with respect to Bob's local reference frame due to Bob's lack of knowledge about the relative rotation that relates his local RF to Alice's RF, i.e. $\rho_{B}=\G[\rho]$. As explained in the previous section this decoherence effect can be taken into account by analysing the efficiency of communication in the presence of noise, i.e. we can assume that Alice and Bob have access to a shared CRF but they only have access to a noisy quantum channel as their means of communication, as depicted in figure $3b$. The type of noisy channel is dictated by the type of reference frame that the parties do not share. For instance, in the case of Cartesian reference frame, the generators $\hat{G}$ are the generators of the group SO(3). Motivated by the analysis of the previous section, again we split the problem into two subproblems.\\

The first case is the case of commutative noise, i.e. when $\hat{K}$ and $\hat{G}$ commute. In this case the noisy quantum channel commutes with the unitary encoding process, i.e. $\G[U_{\lambda}|\psi_{\lambda}\rangle\langle\psi_{\lambda}|U_{\lambda}^{\dag}]=U_{\lambda}\G[|\psi_{\lambda}\rangle\langle\psi_{\lambda}|]U_{\lambda}^{\dag}$ for every $\lambda$. While this property prevents the parties to be able to communicate with encoding the message solely in the qubit, sending a sample of Alice's local CRF together with the qubit makes the communication scheme plausible~\cite{CQC}. As mentioned earlier, this is due to the existence of DFSs. In fact in such cases, Alice has access to the states that remain invariant under the noisy channel $\G$, i.e. states for which we have \footnote{This is the definition of $\G$-invariant states.}$\G[|\psi_{\lambda}\rangle\langle\psi_{\lambda}|]=|\psi_{\lambda}\rangle\langle\psi_{\lambda}|$. As an example consider the case where Alice and Bob do not share a phase reference frame. If Alice encodes $\lambda$ only using a single harmonic oscillator, then the states that she can prepare are of the general form $\sum_{n}c_{n}|n\rangle$ and the operator $\hat{G}$ is the number operator $\hat{N}$. It is easy to check that in this case these states get completely decohered from Bob's point of view. In contrast if Alice chooses two quantum harmonic oscillators as the carrier of her message and the operator $\hat{G}=\hat{N}\otimes\I+\I\otimes\hat{N}$, then the states of the form $a\ket{0,n}+b\ket{1,n-1}$ are invariant under the action of the g-twirling map. For an explicit example of this case we refer the readers to the first example given in section~\ref{Ex}.\\

 The second case is the case of non-commutative noise, i.e. when the operators $\hat{K}$ and $\hat{G}$ do not commute. This case is particularly interesting since estimation of the parameter is possible even in the absence of DFSs as explained in section~\ref{correlation}. In the second and third example of section~\ref{Ex}, we present two scenarios in which the absence of an ideal CRF results in non-commutative noise.\\

\begin{figure}[t]
\includegraphics[width=\linewidth]{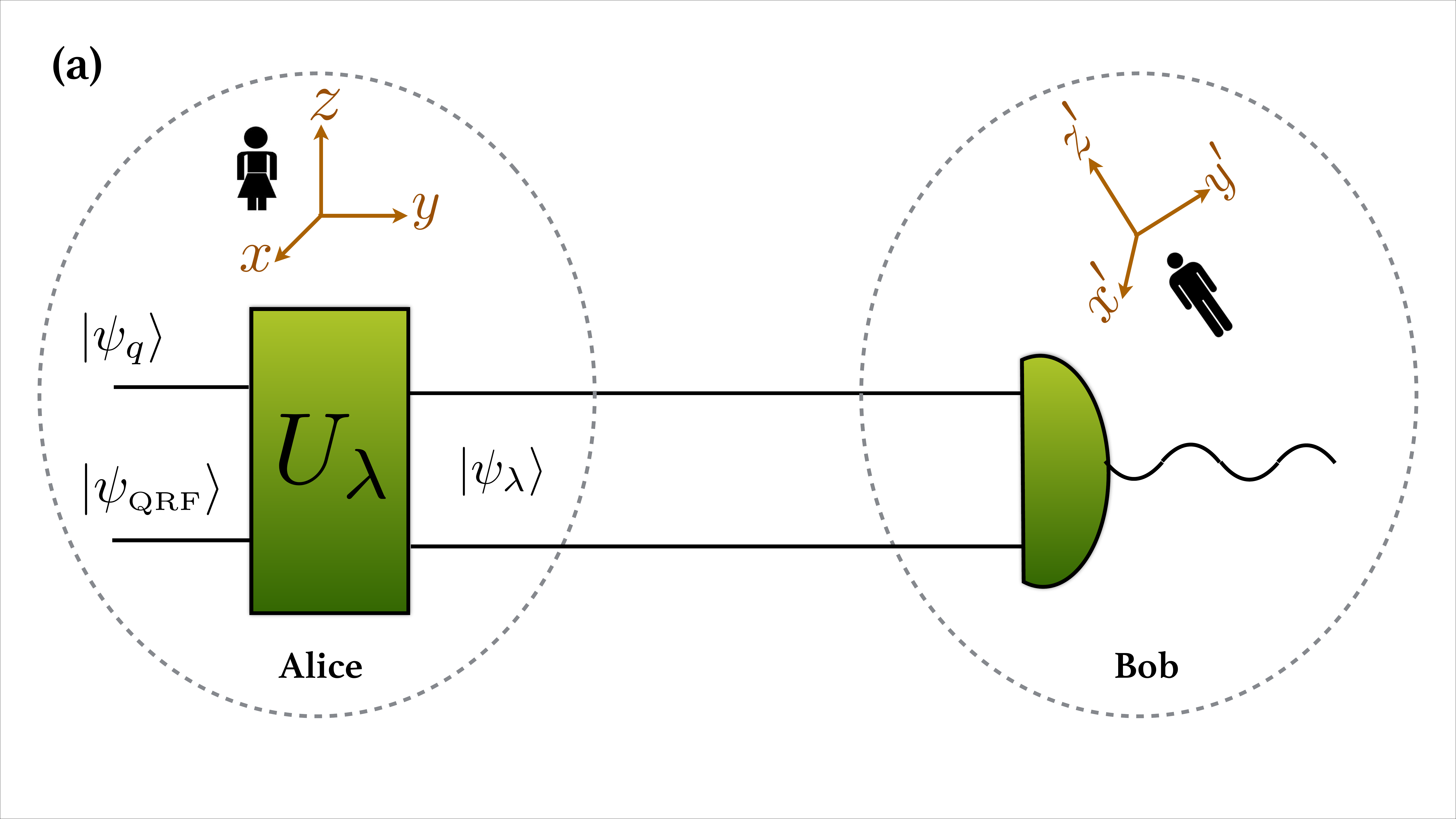}
\includegraphics[width=\linewidth]{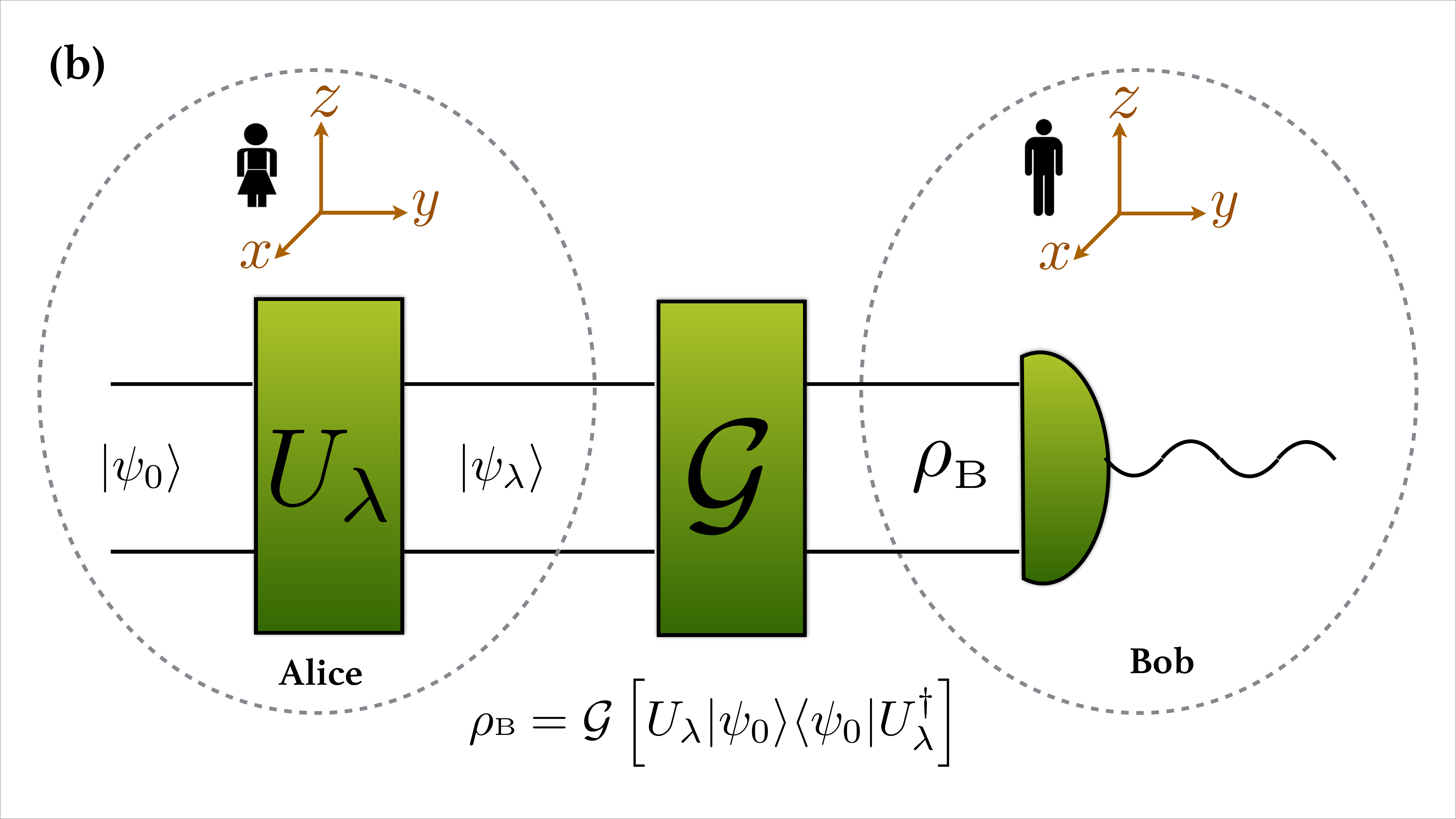}
\caption{a) Communication between two parties in the absence of aligned classical reference frames. b) The effect of misalignment can be viewed as a noisy channel $\G$ in the presence of a shared CRF between Alice and Bob.}
\end{figure}

\subsection{Examples}\label{Ex}
In the previous sections we analysed how QRFs modify our precision in measurement of physical parameters such as time, phase, direction in space, etc. We also explained how misalignment of local RFs is connected with commutative and non-commutative noise in quantum parameter estimation. We are now in place to present some explicit examples.

\begin{figure}[t]
\includegraphics[width=\linewidth]{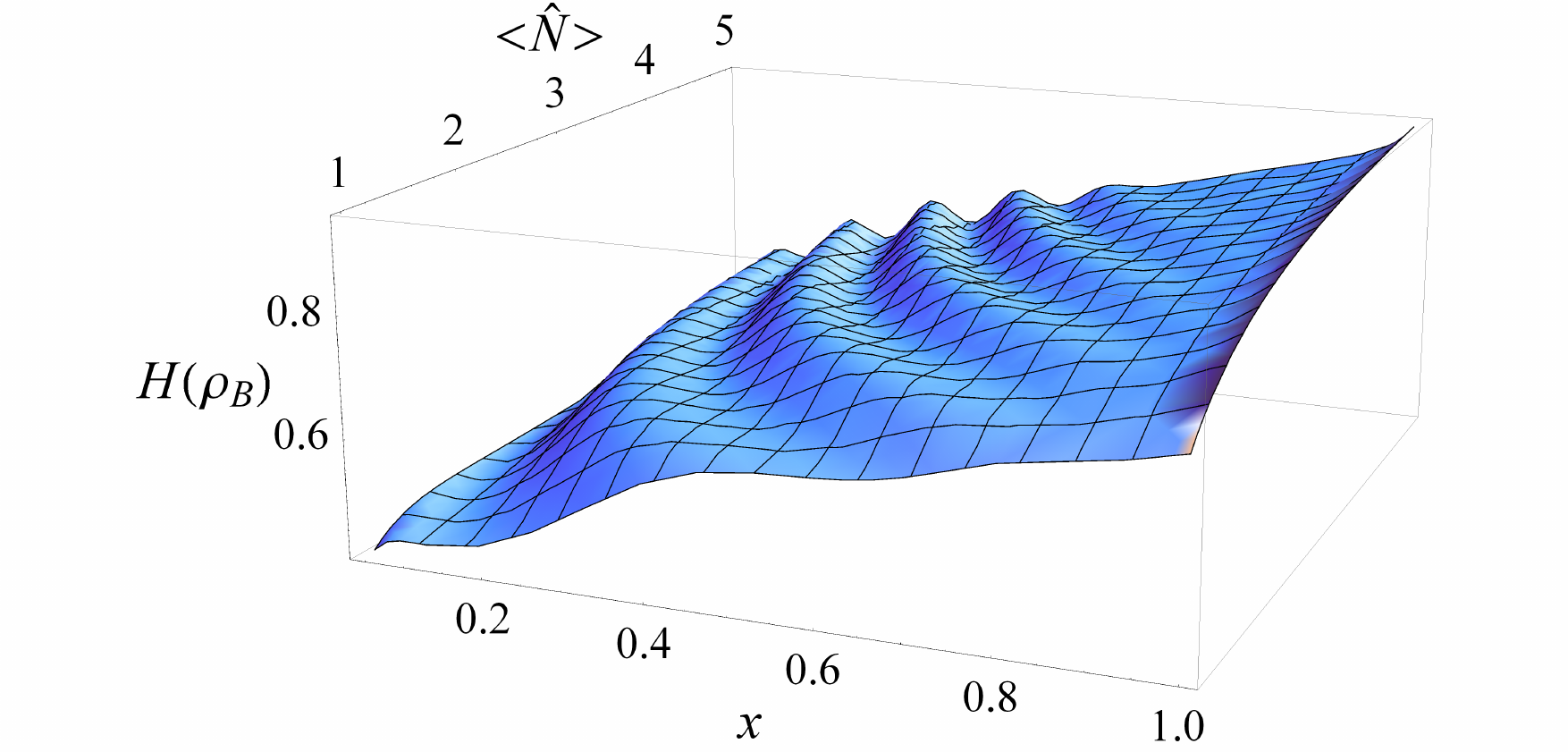}
\caption{Bob's QFI in terms of mean photon number ${\mean{\hat{N}}}$ and $x$ for a squeezed, dispalced vacuum state, i.e. $|\psi_{QRF}\rangle=|\alpha,r\rangle$, as the initial state of the QRF. Paramter $x$
denotes the fraction of mean energy due to displacing the vacuum, i.e. $x=\frac{\alpha^2}{\mean{\hat{N}}}$.}\label{SQD}
\end{figure}
\subsubsection{Example (I): Two non-interacting quantum harmonic oscillators}
The scenario that we consider in this example is as follows. Alice and Bob do not have access to synchronised clocks, i.e. they do not share a common classical RF for time. Alice prepares a state $|\psi_{\lambda}\rangle= U_{\lambda}|\psi_{0}\rangle$, where $U_{\lambda}=e^{i \hat{K}\lambda}$ and $\hat{K}$ is the operator which imprints the parameter $\lambda$ into the fiducial state $|\psi_{0}\rangle$. Since the local clocks of the parties are not synchronised, in Bob's frame the state of the system is given by Eq.~\eqref{gtwirl}, where $U(t)=e^{-i\hat{H}t}$ and $\hat{G}\equiv\hat{H}$ is the Hamiltonian of the qubit and the $QRF$. The operators $\hat{P}_{i}$ are the projectors into subspaces with total energy $E_{i}$. We analyse the quantum Fisher information of the state $\rho_{B}=\G[\rho]$ which tells us how precise Bob will be able to measure $\lambda$.\\

Let us consider the example of two non-interacting quantum harmonic oscillators with the Hamiltonian $H=\hbar \omega (a^{\dag} a+b^{\dag}b)$. The fiducial state is of the product form $|\psi_0\rangle=|\psi_q\rangle \otimes |\psi_{QRF}\rangle$, where $|\psi_q\rangle=(|0\rangle+|1\rangle)/\sqrt{2}$ and $|0\rangle$ and $|1\rangle$ are the eigenstates of number operator $\hat{N_q}=a^{\dag}a$ with eigenvalues $0$ and $1$ respectively. We choose the generator of the unitary channel $U_{\lambda}$ to be $\hat{K}=a^{\dag}a$. It is worth emphasising at this point that in this example $[\hat{K},\hat{H}]=0$. Note that this example is similar to the quantum communication scheme between two parties when they do not have a common phase reference frame as was considered in~\cite{QCQRF}.\\

Using Eq.~\eqref{QFIunitary}, it is straightforward to find the QFI in Alice's frame as $H(\rho)=1$. Note that Alice's QFI is independent of the state of the QRF. On the other hand, if we consider the state $|\psi_{QRF}\rangle=\sum_{n=0}^{N-1}c_{n}|n\rangle$, then using either Eq.~\eqref{commutingKG} or Eq.~\eqref{HrhoB}, we find the QFI in Bob's frame as
\[\label{foranystate}
H(\rho_B)=2-2\left(\sum_{n=0}^{N-2}\frac{\abs{c_n}^4}{\abs{c_n}^2+\abs{c_{n+1}}^2}+\abs{c_{N-1}}^2\right).
\]
If Alice chooses a uniform superposition of Fock states, i.e. the state $|\psi_{US}\rangle=\frac{1}{\sqrt{N}}\sum_{n=0}^{N-1}|n\rangle$, then using~\eqref{foranystate} we can easily compute Bob's QFI as $1-\frac{1}{N}$. Using Eq.~\eqref{SLD}, we find the elusive SLD for this case as
\[
L(\rho_{B,US})=\sum_{n=1}^{N-1}ie^{i\lambda}\ket{0}\pro{n}{n-1}\bra{1}-ie^{-i\lambda}\ket{1}\pro{n-1}{n}\bra{0}.
\]

\begin{figure}[t]
\includegraphics[width=\linewidth]{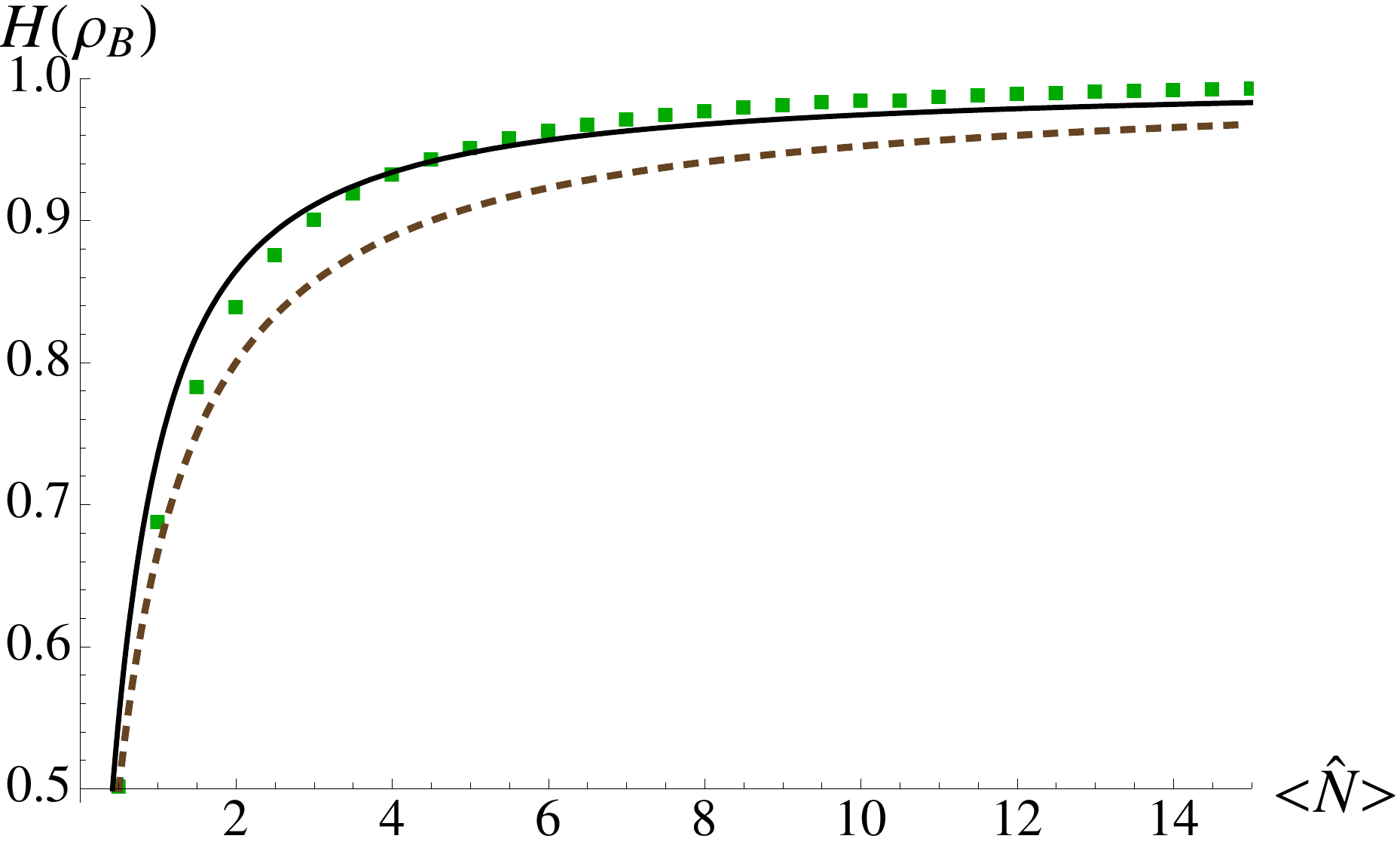}
\caption{Bob's QFI in terms of mean photon number in the initial state of the QRF for three different states. The solid-black, dashed-brown and dotted-green curves correspond to coherent state, uniform superposition state $|\psi_{\tiny{US}}\rangle$, and the optimal state.}\label{comp}
\end{figure}
The SLD provides us with the optimal observable to measure in order to minimise the statistical error in measurement of $\lambda$ and saturate the quantum Cram\'{e}r-Rao bound. This can be easily verified by checking that  $L(\rho_{B,US})$ satisfies the condition~\eqref{HSLDrelation}.

Let us next consider a squeezed, displaced vacuum state~\cite{Gerry} as the state of the QRF, i.e.
\[\label{SD}
|\alpha,r\rangle = \frac{e^{-\frac{\alpha^2}{2}(1+\tanh r)}}{\sqrt{ \cosh r}}
\sum_{n=0}^{\infty}\frac{(\tanh r)^{\frac{n}{2}}}{\sqrt{ 2^{n} n!}}H_{n}\left(\frac{\gamma}{\sqrt{\sinh 2r}}\right)\ket{n}
\]

where $H_{n}(x)$ is the Hermite polynomial, $r$ is the squeezing parameter, $\alpha$ is the displacement parameter and $\gamma=\alpha \text{exp}(r)$. The mean energy of this state is equal to $\alpha^2+\sinh^{2}r$. We define parameter $x$ as the fraction of initial mean energy due to displacing the vacuum, i.e. $x=\frac{\alpha^2}{\mean{\hat{N}}}$. Note that with this definition, $x=0$ and $x=1$ represent a squeezed state and a coherent state respectively. In particular, noticing that in the Fock basis a squeezed state is of the form $|r\rangle=\sum_{n}c_{n}|2n\rangle$ together with Eq.~\eqref{gtwirl}, we find that $H(\G[|\psi_q,r\rangle\langle r,\psi_q|])=0$, i.e. Bob won't be able to decode $\lambda$ if Alice prepares the QRF in a squeezed state.\\

In figure \ref{SQD} we have plotted Bob's QFI for the state $|\alpha,\xi\rangle$ in terms of $x$ and $\mean{\hat{N}}$. As can be seen in this figure, if we fix the mean energy of the QRF, then it is optimal to have zero squeezing in the initial state of the QRF, i.e. $x=1$. This corresponds to preparing the QRF in a coherent state. Using Eq.~\eqref{foranystate} we find Bob's QFI for a coherent state as
 \begin{equation}\label{CohQFI}
H(\rho_B)=2\frac{\abs{\alpha}^2}{1+\abs{\alpha}^2}M\left(1,2+\abs{\alpha}^2,-\abs{\alpha}^2\right),
\end{equation}
where $M(a,b,z)$ is a Confluent hypergeometric function. We derive the asymptotic expression for the limit of large
mean energy, i.e. $\alpha\rightarrow\infty$, as

\[\label{CoherentH}
H(\rho_B)\asymp1-\frac{C}{\abs{\alpha}^2+1},
\]
where $0.25\leq C\leq 0.250001$, i.e. $C\asymp\frac{1}{4}$ (See appendix \ref{appendixCoh} for details).\footnote{For $\abs{\alpha}^2=6$ we already have relative error smaller than $0.01$.}\\

In figure \ref{comp}, we compare Bob's QFI for different states chosen by Alice as a quantum sample of her local RF. This figure shows that a coherent state outperforms the uniform superposition of Fock states. This is in complete agreement with the results of~\cite{WAY} where it is shown that if Bob chooses the Maximum-likelihood estimation process to decode $\lambda$, then choosing a coherent state as the initial state of the QRF instead of the state $|\psi_{US}\rangle$
improves the efficiency of the communication protocol.\\

Also we maximise the QFI in \eqref{foranystate} numerically, which  provides us with the probability amplitudes of the optimal state for fixed $N$, i.e. the state that maximises the QFI or minimises Bob's statistical error in measuring $\lambda$. The green square-shaped dots in figure \ref{comp} represent the QFI for the optimal state. As can be seen from the figure the coherent state is nearly optimal in this case.

\begin{figure}[t]
\includegraphics[width=\linewidth]{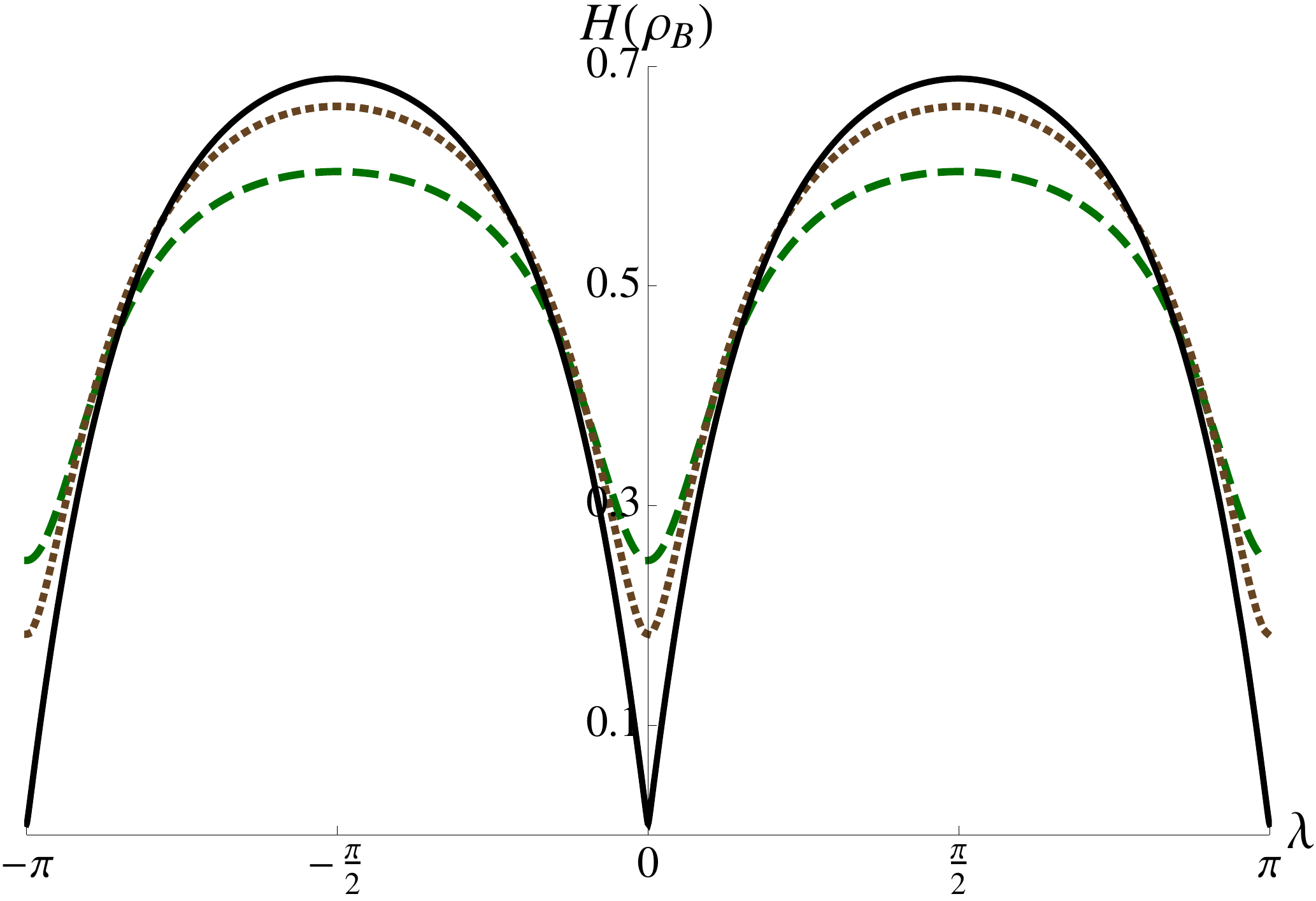}
\caption{Bob's QFI vs. $\lambda$ for two interacting quantum harmonic oscillators. The initial state is considered as $|\psi_{0}\rangle= \frac{1}{\sqrt{2}}(|0\rangle+|1\rangle)\otimes|\psi_{\tiny{US}}\rangle$. The dashed(green), dotted(brown) and solid(black) curves correspond to $N=4,10$ and $300$ respectively.}\label{nonvsint}
\end{figure}

\subsubsection{Example (II): Two interacting quantum harmonic oscillators}
The authors of~\cite{PM} considered  a system of two non-interacting harmonic oscillators. They showed that if one of the harmonic oscillators is used as a quantum clock for the other one, the resultant dynamics will be an approximation to Schr\"{o}dinger dynamics. In this section we investigate the quality of such quantum clocks when the system under study and the clock interact with each other. We analyse the accuracy with which the phase of a quantum system can be measured when we don't have access to an ideal classical clock.\\

Let us consider the example of two interacting quantum harmonic oscillators with the total Hamiltonian
\[
\hat{H}=\hbar\omega(a^{\dagger}a+b^{\dagger}b)+\hbar \kappa(a^{\dagger}b+b^{\dagger}a),
\]
where $\kappa$ is the interaction strength. Similar to the example of two non-interacting quantum harmonic oscillators, we consider the generator of the unitary channel to be the number operator, i.e. $\hat{K}=a^{\dag}a$. Note that the two operators $\hat{K}$ and $\hat{H}$ do not commute in this case, $[\hat{K},\hat{H}]=\kappa (a^{\dag}b-ab^{\dag})$. As mentioned earlier whenever these two operators do not commute, even in the absence of degenerate subspaces of total energy, we may still be able to estimate the parameter. For simplicity we assume that frequency $\omega$ is not a fraction of the interaction strength $\kappa$, i.e.\
\[
\forall P,R\in \mathbb{Z},\ P\omega\neq R\kappa.
\]

This assumption ensures that the hamiltonian $\hat{H}$ does not possess any degenerate eigenvalues.  In order to make the computations easier, we change of the basis as \cite{Twocoupled}
\[
A=\frac{1}{\sqrt{2}}(a+b),\ \ B=\frac{1}{\sqrt{2}}(a-b).
\]
This change of basis allows us to write the Hamiltonian as
$\hat{H}=\hbar(\omega+\kappa)A^{\dagger}A+\hbar(\omega-\kappa)B^{\dagger}B$
with the eigenvectors
\[
\ket{\widetilde{m,n}}\!=\!\frac{(A^{\dagger})^m}{\sqrt{m!}}\frac{(B^{\dagger})^n}{\sqrt{n!}}\ket{\widetilde{0,0}}\!,
\]
which can be written in terms of the Fock basis as
\begin{multline}
\ket{\widetilde{m,n}}=\sum_{k=0}^m\sum_{l=0}^n{{m}\choose{k}}{{n}\choose{l}}\sqrt{\frac{(k+l)!(m+n-k-l)!}{2^{m+n}m!n!}}\\
\ket{k+l,m+n-k-l}.
\end{multline}

\begin{figure}[t]
\includegraphics[width=0.30\textwidth]{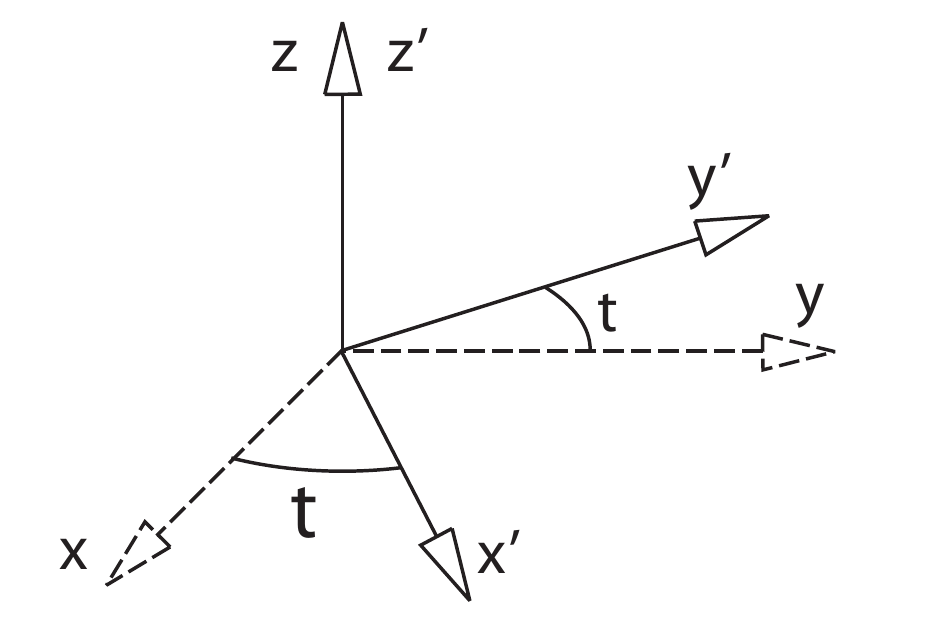}
\caption{Bob only shares his z-axis with Alice, i.e. he is lacks the knowledge about the angle $t$ that related his other two axes to Alice's.}\label{BobRotated}
\end{figure}

Now let us consider that the QRF is initially prepared in the uniform superposition of Fock states.  Using Eq.~\eqref{HrhoB}, we derive the QFI of the averaged state as
\begin{multline}
H(\rho_B)=2-\frac{8}{N}\Bigg(\sum_{m=1}^{\lfloor\frac{N}{2}\rfloor}\sum_{n=0}^{m-1}c_{m,n}(1-d_{m,n}(\lambda))\\
-\!\!\!\!\sum_{m=\lfloor\frac{N}{2}\rfloor+1}^{N}\!\sum_{n=0}^{N-m}c_{m,n}-\!\!\!\!\sum_{m=\lfloor\frac{N}{2}\rfloor+1}^{N-1}\!\!\!\!\sum_{n=0}^{N-m-1}c_{m,n}d_{m,n}(\lambda)\Bigg),
\end{multline}
where $d_{m,n}(\lambda)=\frac{(m+n)((m-n)^2+m+n)\sin^2\lambda}{((m-n)^2-m-n)^2+4(m+n)(m-n)^2\sin^2\lambda}$, \linebreak
$c_{m,n}=\frac{(m+n-1)!(m-n)^2}{2^{m+n+1}m!n!}$ and $\lfloor\cdot\rfloor$ is a floor function.\\

In this example since $\hat{K}$ and $\hat{G}$ do not commute, $H(\G[\rho])$ is $\lambda$-dependent as opposed to the first example where Bob's QFI was independent of the encoded parameter $\lambda$. In figure \ref{nonvsint} we have plotted the QFI $H(\G[\rho])$ in terms of $\lambda$ for increasing values of the mean energy in the state of the QRF. The maximum and minimum of the QFI occurs at $\lambda=\pm \frac{\pi}{2}$ and $\lambda=0,\pm\pi$ respectively. Note that even for very large $N$ QFI does not approach the ideal case. In other words, even in the limit of very large mean energy in the initial state of the quantum clock, we can not estimate the phase parameter $\lambda$ as precise as we could if we had access to a classical clock. This can be proved using necessary conditions~\eqref{noloss}~and~\eqref{maxlossKG}. One can easily check that $\mathrm{Cov}_\rho(\hat{G},\hat{K})=\frac{\hbar\omega}{4}$, which means that independent of $N$ and $\lambda$, the QFI is always smaller than one, i.e. $H(\G[\rho])<1$. Similarly, since $\mean{[\hat{K},\hat{G}]}_\rho\approx\frac{2i\hbar\kappa}{3}\sqrt{N}\sin\lambda$, we can deduce that independent of $N$ for $\lambda\neq-\pi,0,\pi$, the QFI is always positive.

\begin{figure}[t]
\includegraphics[width=\linewidth]{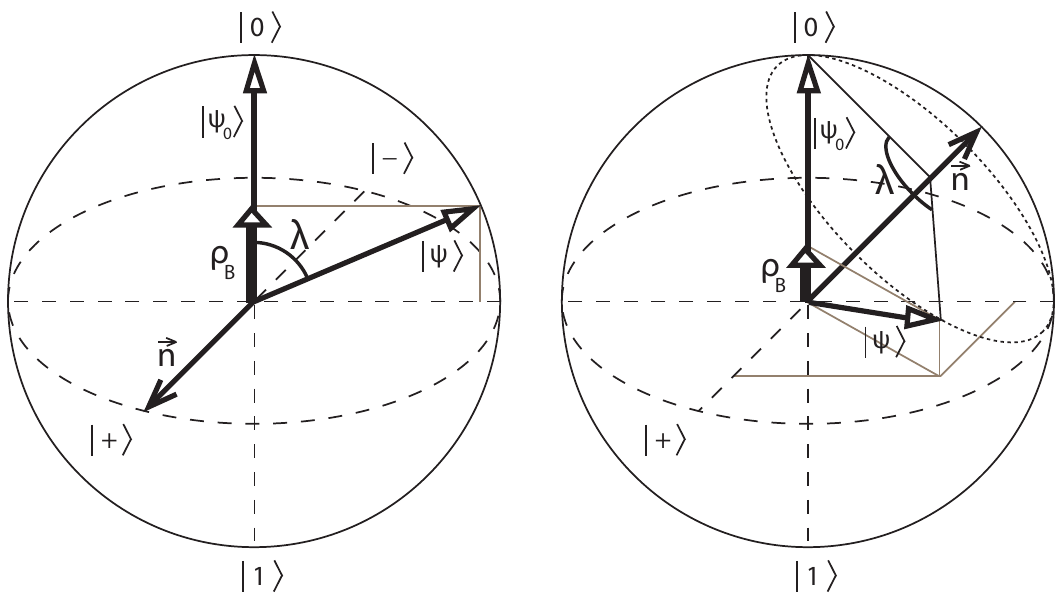}
\caption{Encoding $\lambda$ via rotating the fiducial state $\ket{\psi_0}=\ket{0}$ around the unit vector $\vec{n}$. For $\vec{n}=(1,0,0)$ the state of the qubit in Bob's frame is $\rho_B=\cos^2(\frac{\lambda}{2})\pro{0}{0}+\sin^2(\frac{\lambda}{2})\pro{1}{1}$ and Bob's QFI is the same as Alice's. For $\vec{n}=(0,\frac{1}{\sqrt{2}},\frac{1}{\sqrt{2}})$, $\rho_B$ is $\rho_B=(1-\frac{1}{2}\sin^2(\frac{\lambda}{2}))\pro{0}{0}+\frac{1}{2}\sin^2(\frac{\lambda}{2})\pro{1}{1}$. Note that $\rho_{B}$ is the projection of $|\psi\rangle$ onto the z-axis. Also note that in the latter case Bob's QFI is $\lambda$-dependent(See figure \ref{BobsQFIexample3}).}\label{BlochSpheres}
\end{figure}

\subsubsection{Example (III): Direction indicator}\label{RotatingObserver}
In the first example we observed how using a QRF enables Alice and Bob to perform an alignment-free communication protocol in the presence of commutative noise, while in the second example we analysed how the absence of a perfect CRF can cause non-commutative noise which then reduces the precision with which a physical parameter can be estimated. Here, we present an example in which the noise caused due to Bob's lack of knowledge about Alice's local reference frame is non-commutative. We analyse how precise Bob can extract $\lambda$ if Alice does not send him a  quantum sample of her local RF.\\

Let us start with the case where Alice wishes to both encode and decode a parameter herself. She chooses a spin-$\frac{1}{2}$ particle as the physical system to encode a parameter $\lambda$ and then she encodes this parameter using a unitary channel with the generator
\[
\hat{K}=\frac{1}{2}\vec{n}\cdot\vec{\sigma}=\frac{1}{2}(x\sigma_x+y\sigma_y+z\sigma_z).
\]
This is the generator of a general rotation in the Bloch sphere around the axis $\vec{n}=(x,y,z)$, where $x^2+y^2+z^2=1$ and $x,y,z$ are real parameters.
For simplicity we choose the fiducial state to be the eigenstate of $\sigma_z$ with eigenvalue $1$, i.e. $\ket{\psi_0}=\ket{0}$.
Using Euler's formula for Pauli matrices\footnote{$e^{-i\hat{K}\lambda}=\cos(\frac{\lambda}{2})\mathbb{I}-i\sin(\frac{\lambda}{2})(\vec{n}\cdot\vec{\sigma})$}, we can write Alice's prepared state as
\begin{equation}\label{EX3state}
\ket{\psi_{\lambda}}=\left(\cos\left(\tfrac{\lambda}{2}\right)-iz\sin\left(\tfrac{\lambda}{2}\right)\right)\ket{0}+
(y-ix)\sin\left(\tfrac{\lambda}{2}\right)\ket{1}.
\end{equation}

Then using Eq.~\eqref{QFIunitary}, the QFI in Alice's frame reads as
\[\label{example3AlicesH}
H(\rho)=1-z^2 .
\]
Note that for $z=1$, the corresponding generator is $\hat{K}=\frac{1}{2}\sigma_z$ which leaves the fiducial state invariant, i.e. $\text{exp}(-i\frac{\sigma_{z}}{2})|0\rangle=|0\rangle$. Since the encoding process is not successful, the QFI, $H(\rho)$, vanishes which simply means that a different generator needs to be used at the preparation stage. The QFI takes its maximum value when when the parameter $\lambda$ is encoded via a rotation around any vector in the $xy$-plane, i.e. when $z=0$.\\

\begin{figure}[t]
\includegraphics[width=\linewidth]{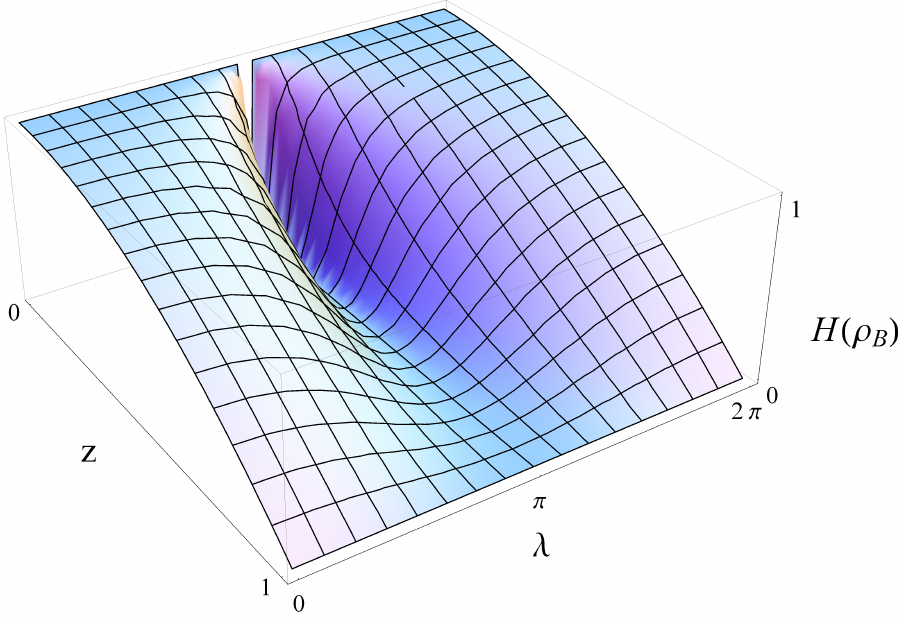}
\caption{Bob's QFI in terms of $\lambda$ and $z$ for general $\vec{n}=(x,y,z)$.}\label{BobsQFIexample3}
\end{figure}

Now suppose that Alice and Bob only share their $z$-axis, i.e. Bob is completely unaware of the relative angle $t$ between his other two axes and Alice's, as depicted in figure \ref{BobRotated}. In this case, $\hat{G}$ is the generator of rotations around $z$-axis, i.e. $\hat{G}=\frac{1}{2}\sigma_z$. Using Eq.~\eqref{noncommutingKG}, the QFI in Bob's frame can be written as
\[\label{example3BobsH}
H(\rho_B)=\frac{1-z^2}{1+z^2\tan^2\left(\tfrac{\lambda}{2}\right)}=\frac{H(\rho)}{1+z^2\tan^2\left(\tfrac{\lambda}{2}\right)}.
\]

Again note that for $z=1$, the QFI is zero in Bob's frame. This is expected, since Bob lacks some information with respect to Alice, therefore Alice's  inability in extracting information about $\lambda$ means that Bob will not be able to decode the message either, i.e. $H(\rho_{B})=0$. On the other hand, as can be seen from \eqref{example3BobsH}, when $z=0$ the QFI is the same in Alice's frame and Bob's frame. Figure \ref{BlochSpheres} depicts the two cases of $\vec{n}=(1,0,0)$ and $\vec{n}=(0,\frac{1}{\sqrt{2}},\frac{1}{\sqrt{2}})$. For the former case, the efficiency of communication is $\lambda$-independent, whereas for the latter case it is $\lambda$-dependent, as can be seen in figure \ref{BobsQFIexample3}. In this figure, we have plotted Bob's QFI in terms of $\lambda$ and $z$ for general $\vec{n}=(x,y,z)$. We observe that as $ \lambda$ approaches the value $\pi$, the QFI approaches its minimum value, i.e.
$H(\rho_B)\rightarrow 0$. In other words, for the chosen encoding operator $\hat{K}$ and the fiducial state $|0\rangle$, Bob will not be able to distinguish $\rho_{\pi}$ form its neighbouring states $\rho_{\pi\pm\epsilon}$, where $\epsilon$ is a very small change in $\lambda=\pi$.\\

Also after some algebra and with the aid of Eq.~\eqref{NSLD}, we find the SLD operator that achieves the QFI in~\eqref{example3BobsH} as\\
\[
L(\rho_B)=\frac{(z^2-1)\tan\left(\tfrac{\lambda}{2}\right)}{1+z^2\tan^2\left(\tfrac{\lambda}{2}\right)}\pro{0}{0}+\cot\left(\tfrac{\lambda}{2}\right)\pro{1}{1}.
\]
Again the optimal POVM can be constructed from the eigenvalues of this operator, i.e. $\{\ket{0}\langle 0|,\ket{1}\langle 1|\}$. This simply means that the most informative measurement for Bob is the measurement in the computational basis. In order to verify  that this is in fact the case, we can either use the  relation~\eqref{HSLDrelation} or we can compute the classical Fisher information using Eq.~\eqref{CFI} and show that it is equal to the QFI~\eqref{example3BobsH}.


\section{Discussions and outlook}\label{DO}
In quantum metrological schemes the existence of a prefect classical reference frame is often assumed. Here we have exploited the powerful mathematical tools from quantum metrology in order to analyse the modification of the ultimate precision limits due to the absence of such frames of reference. We considered the effects of commutative and non-commutative noise due to lack of a certain CRF. In doing so, we showed that the more the encoding process and the nature of the noise resemble each other, the more precision is lost. We also presented necessary and sufficient conditions for two extreme cases. The first case is when the absence of an ideal RF does not reduce the accuracy of estimation and the second case is when the estimation of the parameter with respect to QRFs is no longer possible. Moreover, by explaining the connection between noisy parameter estimation protocols and alignment-free communication schemes~\cite{SSR-RF}, we shed light into different aspects of quantum communication in the absence of aligned reference frames.\\

Our future line of research includes incorporating other sources of noise in the alignment-free communication protocols. We are interested in the regimes where relativity starts to play a more significant role~\cite{Alphacen}. Relativistic effects such as the decoherence caused due to non-uniform motion~\cite{tele} or the effects of the gravitational field of the earth~\cite{DavTimIvette} are the possible sources of noise that yet need to be considered. Recently in \cite{RQM1,RQM2} techniques for the optimal estimation of parameters which appear in quantum field theory in curved spacetime have been presented. This enables the estimation of parameters such as proper acceleration, proper time, relative distance, amplitude of gravitational waves~\cite{GW}, as well as spacetime parameters of interest, such as the expansion rate of the Universe or the mass of a black hole. Finally, the reference frames of  relativistic observers is inevitably misaligned with respect to non-relativistic observers, therefore it is crucial to consider the effect of relativistic noise on estimation of parameters of interest such as acceleration, time, phase, temperature, etc.\\

\emph{Acknowledgements}:\
We thank Antony Lee, Katarzyna Macieszczak and  Carlos Sab{\'\i}n for useful discussions and comments. M.~A. and I.~F. acknowledge support from EPSRC (CAF Grant No.~EP/G00496X/2 to I.~F.).\\

\appendix
\section{Derivation of QFI and the SLD operator in the absence of perfect RFs}\label{FirstFormula}
From Eq.~\eqref{rhoB}, we immediately observe that eigenvalues of transformed density matrix $\rho_B$ are $p_i=\langle\psi_{\lambda}|\hat{P}_i|\psi_{\lambda}\rangle$ with respective normalised eigenvectors $\frac{\hat{P_i}\ket{\psi_{\lambda}}}{\sqrt{p_i}}$. Let $\{\ket{\phi_j}\}_j$ be a set orthonormal eigenvectors of $\rho_B$ with respective eigenvalue $0$. 
Using Eq.~(\ref{QFI}) we have
\begin{equation}
\begin{split}
H(\rho_B)&=\ \ \ 2\!\!\!\!\!\!\!\!\!\!\sum_{i,j,p_i\neq0,p_j\neq0}\!\!\!\!\frac{\big|\frac{\bra{\psi}\hat{P_i}}{\sqrt{p_i}}\sum_k\hat{P}_k\partial_\lambda\rho \hat{P}_k\frac{\hat{P}_j\ket{\psi}}{\sqrt{p_j}}\big|^2}{p_i+p_j}\\
&\ \ \ \ \!+4\sum_{i,j}\frac{\big|\frac{\bra{\psi}\hat{P_i}}{\sqrt{p_i}}\sum_k \hat{P}_k\partial_\lambda\rho \hat{P}_k\ket{\phi_j}\big|^2}{p_i}\\
&=\sum_i\frac{\big|\frac{\bra{\psi}\hat{P_i}}{\sqrt{p_i}}\partial_\lambda\rho \frac{\hat{P_i}\ket{\psi}}{\sqrt{p_j}}\big|^2+4\sum_j\big|\frac{\bra{\psi}\hat{P_i}}{\sqrt{p_i}}\partial_\lambda\rho \hat{P_i}\ket{\phi_j}\big|^2}{p_i}
\end{split}
\end{equation}
and together with the Parseval identity, i.e.
\begin{equation}
\begin{split}
\sum_j\Big|\frac{\bra{\psi}\hat{P}_i}{\sqrt{p_i}}\partial_\lambda\rho \hat{P}_i\ket{\phi_j}\Big|^2&=
\Big|\!\Big|\hat{P}_i\partial_\lambda\rho\frac{\hat{P}_i\ket{\psi}}{\sqrt{p_i}}\Big|\!\Big|^2\\
&-\sum_j\Big|\frac{\bra{\psi}\hat{P}_i}{\sqrt{p_i}}\partial_\lambda\rho \hat{P}_i\frac{\hat{P}_j\ket{\psi}}{\sqrt{p_j}}\Big|^2
\end{split}
\end{equation}
we can remove the dependence on states $\ket{\phi_j}$. Then $H(\rho_{B})$ is
\[
H(\rho_B)\!=\!\!\sum_i\!\!\frac{4p_i\bra{\psi}\hat{P}_i\partial_\lambda\rho \hat{P}_{i}\partial_\lambda\rho \hat{P}_{i}\ket{\psi}\!-\!3\abs{\bra{\psi}\hat{P}_{i}\partial_\lambda\rho \hat{P}_{i}\ket{\psi}}^2}{p_i^3}.
\]
After substituting $\rho_{\lambda}=\pro{\psi_{\lambda}}{\psi_{\lambda}}$
\[\label{prefinal}
H(\rho_B)\!=\!\!\!\!\sum_{i,p_i\neq0}\!\!\!4\bra{\partial_\lambda\psi}\hat{P}_{i}\ket{\partial_\lambda\psi}+
\frac{(\bra{\psi}\hat{P}_{i}\ket{\partial_\lambda\psi}\!-\!\bra{\partial_\lambda\psi}\hat{P}_{i}\ket{\psi})^2}{p_i},
\]
where $\ket{\partial_\lambda\psi}=\sum_k(\partial_\lambda\alpha_k)\ket{k}$ for $\lambda$-independent basis $\{\ket{k}\}$. The sum in (\ref{prefinal}) consists only of elements where $p_i\neq0$, however, by differentiating $p_i=\bra{\psi}\hat{P}_{i}\ket{\psi}=0$ and using Schwarz inequality on $\bra{\partial_{\lambda\lambda}\psi}\hat{P}_{i}\ket{\psi}$ we get $\bra{\partial_\lambda\psi}\hat{P}_{i}\ket{\partial_\lambda\psi}=0$. Now summing over all $i$ and using the completeness relation $\sum_i\hat{P}_{i}=\mathbf{1}$ we get
\[\label{almostdone}
H(\rho_B)=4\braket{\partial_\lambda\psi}{\partial_\lambda\psi}-
4\sum_i\frac{(\mathfrak{Im}\bra{\psi}\hat{P}_{i}\ket{\partial_\lambda\psi})^2}{\bra{\psi}\hat{P}_{i}\ket{\psi}}.
\]
\emph{Symmetric logarithmic derivative} \eqref{NSLD} can be derived analogously, where instead of Parseval identity we use completeness relation
$\sum_j\pro{\phi_j}{\phi_j}=\mathbb{I}-\sum_i\frac{P_i\pro{\psi}{\psi}P_i}{p_i}$.

\section{Proof of theorem \ref{theorem}}\label{nonnegQFIl}

Here we prove that $0\leq l(\rho,G)\leq H(\rho)$ and the equality conditions. $l(\rho,G)\leq H(\rho)$ follows immediately from definition \eqref{QFIlossdef}. Let us prove $l(\rho,G)\geq 0$. Looking at the expression for QFI loss, i.e. Eq.~\eqref{QFIlossdef}, we need to prove that
\[\label{lnonnegativestart}
\sum_i\frac{(\mathfrak{Im}\bra{\psi}\hat{P}_{i}\ket{\partial_\lambda\psi})^2}{\bra{\psi}\hat{P}_{i}\ket{\psi}}\geq|\braket{\psi}{\partial_\lambda\psi}|^2.
\]
First, let us define $\ket{\widetilde{\partial_\lambda\psi}}:=\sum_i\frac{\mathfrak{Im}\bra{\psi}\hat{P}_{i}\ket{\partial_\lambda\psi}}{\PI}\hat{P}_{i}\ket{\psi}$.
Then using the fact that the state $|\psi\rangle$ is normalised, i.e. $\norm{\ket{\psi}}=1$, the Schwarz inequality and that for any state $|\psi\rangle$, $p_i\equiv\bra{\psi}\hat{P}_{i}\ket{\psi}=0$ if and only if  $\hat{P}_{i}\ket{\psi}=0$ and therefore $\bra{\psi}\hat{P}_{i}\ket{\partial_\lambda\psi}=0$, together with the completeness relation $\sum_i\hat{P}_i=\mathbf{1}$, we have
\begin{equation}
\begin{split}
LHS&=\norm{\ket{\widetilde{\partial_\lambda\psi}}}^2=\norm{\ket{\widetilde{\partial_\lambda\psi}}}^2\norm{\ket{\psi}}^2\geq\abs{\braket{\psi}{\widetilde{\partial_\lambda\psi}}}^2\\
&=\bigg|\bra{\psi}\sum_{i,p_i\neq0}\frac{\mathfrak{Im}\bra{\psi}\hat{P}_{i}\ket{\partial_\lambda\psi}}{\PI}\hat{P}_{i}\ket{\psi}\bigg|^2\\
&=\Big|\mathfrak{Im}\bra{\psi}\sum_i\hat{P}_{i}\ket{\partial_\lambda\psi}\Big|^2=\abs{\mathfrak{Im}\braket{\psi}{\partial_\lambda\psi}}^2\\
&=\left|\braket{\psi}{\partial_\lambda\psi}\right|^2
\end{split}
\end{equation}
where for the last step we have used Eq.~\eqref{purelyimaginary}. Now, because Schwarz inequality is saturated if and only if there exists a complex number $c$ such that $\ket{\widetilde{\partial_\lambda\psi}}=c\ket{\psi}$, by re-writing the state $\ket{\psi}$ as $\sum_i\frac{p_i}{p_i}\hat{P}_{i}\ket{\psi}$ we find that Eq.~\eqref{lnonnegativestart} is saturated if and only if
\[
\sum_i\frac{\mathfrak{Im}\bra{\psi}\hat{P}_{i}\ket{\partial_\lambda\psi}-cp_i}{p_i}\hat{P}_{i}\ket{\psi}=0,
\]
which together with orthogonality condition for the projectors $\hat{P}_{i}$ leads to the no-loss condition \eqref{cnoloss}, i.e.
\[
l(\rho,G)=0\ \Leftrightarrow\ \exists c\in\mathbb{C},\ \forall i,\ \mathfrak{Im}\bra{\psi}\hat{P}_{i}\ket{\partial_\lambda\psi}=c\PI.
\]
Now let us derive the max-loss condition~\eqref{maxlosstheorem}. From the definition of Quantum Fisher information loss and that $\abs{\braket{\psi}{\widetilde{\partial_\lambda\psi}}}=\abs{\braket{\psi}{\partial_\lambda\psi}}$, we can write
\begin{equation}\label{lossformula}
\begin{split}
l(\rho,G)&=4\braket{\widetilde{\partial_\lambda\psi}}{\widetilde{\partial_\lambda\psi}}-4\abs{\braket{\psi}{\widetilde{\partial_\lambda\psi}}}^2\\
&=4\braket{\widetilde{\partial_\lambda\psi}}{\widetilde{\partial_\lambda\psi}}-4\abs{\braket{\psi}{{\partial_\lambda\psi}}}^2.
\end{split}
\end{equation}
Therefore by comparing (\ref{lossformula}) and (\ref{paris}) we have
\[\label{nolossformulaappendix}
l(\rho,\hat{G})=H(\rho)\ \Leftrightarrow
\braket{\widetilde{\partial_\lambda\psi}}{\widetilde{\partial_\lambda\psi}}=\braket{{\partial_\lambda\psi}}{{\partial_\lambda\psi}}.
\]
Similar to the previous case we can write $\ket{\partial_\lambda\psi}$ in the complete orthonormal basis $\{\frac{\hat{P}_{i}\ket{\psi}}{\sqrt{p_i}},\ket{\phi_j}\}_{i,j}$ as
\[
\ket{\partial_\lambda\psi}=\sum_i\frac{\bra{\psi}\hat{P}_{i}\ket{\partial_\lambda\psi}}{\sqrt{p_i}}\frac{\hat{P}_{i}\ket{\psi}}{\sqrt{p_i}}
+\sum_j\braket{\phi_j}{\partial_\lambda\psi}\ket{\phi_j}.
\]
where $\ket{\phi_j}$ span the rest of the Hilbert space which is not spanned by vectors $\frac{\hat{P}_{i}\ket{\psi}}{\sqrt{p_i}}$.
After multiplying by $\bra{\partial_\lambda\psi}$ we get the Parseval identity, i.e.
\[
\braket{\partial_\lambda\psi}{\partial_\lambda\psi}=\sum_i\frac{\abs{\bra{\psi}\hat{P}_{i}\ket{\partial_\lambda\psi}}^2}{p_i}
+\sum_j\abs{\braket{\phi_j}{\partial_\lambda\psi}}^2.
\]
Comparing this with (\ref{nolossformulaappendix}) we get condition for max-loss as $l(\rho,\hat{G})=H(\rho)\ \Leftrightarrow$
\[
\forall i,\ \mathfrak{Re}\bra{\psi}\hat{P}_{i}\ket{\partial_\lambda\psi}=0\ \wedge\ \forall \ket{\phi_j},\ \braket{\phi_j}{\partial_\lambda\psi}=0.
\]

\section{Alternative description in terms of eigenvectors of $\hat{G}$}\label{alternative}

Here we introduce alternative an description of Quantum Fisher information using eigenvectors of the operator $\hat{G}$. Let $\{\ket{v_{i,j}}\}_j$ be a set of orthonormal eigenvectors of $\hat{G}$ with respective eigenvalue $G_i$. Then the projection operator into the eigenspace corresponding to eigenvalue $G_{i}$ can be written as
\[
\hat{P}_i=\sum_j\pro{v_{i,j}}{v_{i,j}}.
\]
Then we substitute this expression into \eqref{paris} and find the QFI in terms of the eigenvectors $\ket{v_{i,j}}$ as
\[\label{eigenformula}
H(\rho_B)\!=\!4\braket{\partial_\lambda\psi}{\partial_\lambda\psi}\!-\!4\!\sum_i
\!\!\frac{\big(\mathfrak{Im}\big(\sum_j\!\braket{\psi}{v_{i,j}}\braket{v_{i,j}}{\partial_\lambda\psi}\big)\!\big)^2}{\sum_j\abs{\braket{v_{i,j}}{\psi}}^2}
\]
In order to write this expression in a more compact way we define the unnormalised states $\cket{a_i},\cket{b_i}\in\mathbb{C}^{n_i}$,

\[
\cket{a_i}_j:=\braket{v_{i,j}}{\psi},\ \cket{b_i}_j:=\braket{v_{i,j}}{\partial_\lambda\psi},
\]
where $n_i$ denotes the dimension of the subspace corresponding to eigenvalue $G_{i}$. This way we can write Eq.~\eqref{eigenformula} as
\[\label{eigform}
H(\rho_B)=4\bigg(\braket{\partial_\lambda\psi}{\partial_\lambda\psi}-\sum_i
\frac{(\mathfrak{Im}\cbraket{a_i}{b_i})^2}{\norm{\cket{a_i}}^2}\bigg),
\]
where $\cbraket{\cdot}{\cdot}$ is a standard inner product on $\mathbb{C}^{n_i}$.\\

Let us here also prove that when $\hat{K}$ and $\hat{G}$ commute and the spectrum of $\hat{G}$ is non-degenerate, then $H(\rho_B)=0$, i.e. all the information about $\lambda$ is lost due to the noise. This can be done either starting from (\ref{eigenformula}) or by using max-loss condition (\ref{maxloss}). Here we choose the latter. Because $\hat{G}$ has non-degenerate spectrum,  the projectors $\hat{P_i}$ are all rank-one projectors, i.e. $\hat{P}_i=\pro{v_{i}}{v_{i}}$. Also since $\hat{K}$ and $\hat{G}$ commute, then all projectors also commute with $\hat{K}$, i.e. $[\hat{K},\pro{v_{i}}{v_{i}}]=0$, where $\{\ket{v_{i}}\}_i$ is a complete basis. Consider a subset of this basis that do not have any overlap with the state $\ket{\psi}$, i.e. the subset $\{\ket{v_j}\in\{\ket{v_i}\}_i|\braket{v_j}{\psi}=0\}_{j}$. This set replaces $\{\ket{\phi_j}\}_j$ in the max-loss condition~(\ref{maxloss}). Now we have
\[
\bra{v_j}\hat{K}\ket{\psi}=\braket{v_j}{v_j}\bra{v_j}\hat{K}\ket{\psi}=\bra{v_j}\hat{K}\ket{v_j}\braket{v_j}{\psi}=0,
\]
which makes the proof complete.

\begin{figure}[t]
\includegraphics[width=\linewidth]{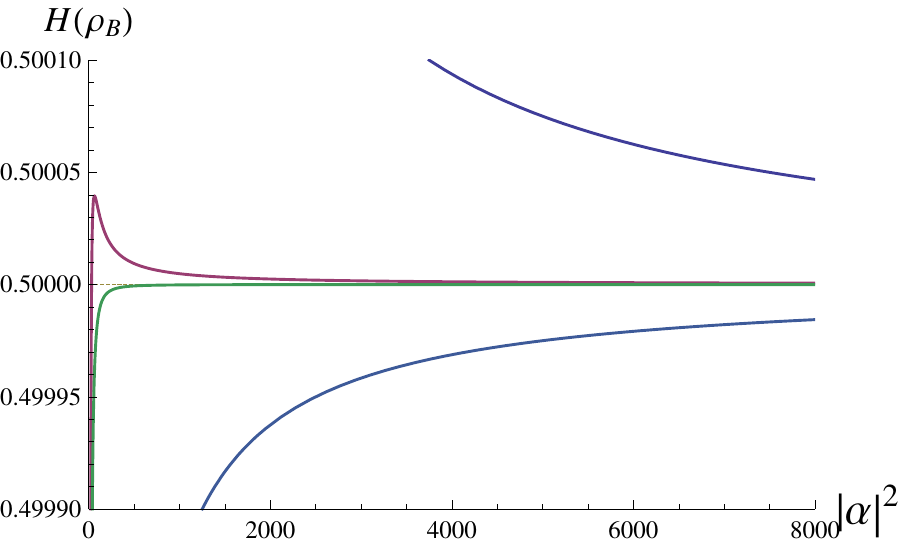}
\caption{Function $\frac{|\alpha|^2}{|\alpha|^2+c}M(|\alpha|^2)$ in terms of $|\alpha|^2$. From top to bottom for $c=0,0.74,0.75$ and $1$.}\label{MCalphaM}
\end{figure}

\section{Asymptotic scaling of $H(\rho_B)$ in example (I) for the coherent state as the initial state of the QRF }\label{appendixCoh}
In the case that the two operators $\hat{K}$ and $\hat{G}$ commute and in the limit of large initial mean energy in the state of the QRF, the classical limit of a quantum reference frame should be recovered, i.e. $H(\rho_B)\rightarrow1$, as was shown in the first example for the uniform superposition of Fock states. Here we analyse this asymptotic behaviour for a coherent state as the initial state of the QRF. We expect
\[
M\left(\abs{\alpha}^2\right):=M\left(1,2+\abs{\alpha}^2,-\abs{\alpha}^2\right)=\tfrac{1}{2}+f\left(\abs{\alpha}^2\right),
\]
such that $\lim_{\abs{\alpha}^2\rightarrow \infty} f(\abs{\alpha}^2)=0$. As figure $\ref{MCalphaM}$ suggests, this is true for large enough mean photon number in the initial state of the coherent state. Moreover, from this figure one can see that, for the mean photon number above a certain threshold, we can find constants $C_1$ and $C_2$ such that
\[
\frac{\abs{\alpha}^2}{C_1+\abs{\alpha}^2}M\left(\abs{\alpha}^2\right)\leq\frac{1}{2} \leq\frac{\abs{\alpha}^2}{C_2+\abs{\alpha}^2}M\left(\abs{\alpha}^2\right).
\]

Re-arranging the expression above, we can find $\alpha$-dependent lower and upper bounds for the function $f$ as
\[
\frac{C_1}{2\abs{\alpha}^2}\leq f\left(\abs{\alpha}^2\right)\leq\frac{C_2}{2\abs{\alpha}^2}.
\]
Therefore the QFI in Bob's reference frame is
\begin{equation}
\begin{split}
H(\rho_B)&=2\frac{\abs{\alpha}^2}{1+\abs{\alpha}^2}\left(\frac{1}{2}+f\left(\abs{\alpha}^2\right)\right)\\
&=1-g\left(\abs{\alpha}^2\right),
\end{split}
\end{equation}
where 
$\frac{1-C_2}{1+\abs{\alpha}^2}\leq g(\abs{\alpha}^2)\leq\frac{1-C_1}{1+\abs{\alpha}^2}$. Choosing $C_1=0.749999$ and $C_2=0.75$, we tighten the lower and upper bounds on $C$, i.e. $0.25\leq C \leq 0.250001$.

\end{document}